# Nanoscale study of reactive transport in catalyst layer of proton exchange membrane fuel cells with precious and non-precious catalysts using lattice Boltzmann method

Li Chen [a, b], Gang Wu [c], Qinjun Kang [b], Edward F Holby [d], Wen-Quan Tao [a]

a: Key Laboratory of Thermo-Fluid Science and Engineering of MOE, School of Energy and Power Engineering, Xi'an Jiaotong University, Xi'an, Shaanxi 710049, China

b: Earth and Environmental Sciences Division, Los Alamos National Laboratory, Los Alamos, New Mexico, USA

c: Materials Physics & Applications Division, Los Alamos National Laboratory, Los Alamos, New Mexico, USA

d: Materials Science & Technology Division, Los Alamos National Laboratory, Los Alamos, New Mexico, USA

**Abstract**

High-resolution porous structures of catalyst layer (CL) with multicomponent in proton exchange membrane fuel cells are reconstructed using a reconstruction method called quartet structure generation set. Characterization analyses of nanoscale structures are implemented including pore size distribution, specific area and phase connectivity. Pore-scale simulation methods based on the lattice Boltzmann method are developed and used to predict the macroscopic transport properties including effective diffusivity and proton conductivity. Nonuniform distributions of ionomer in CL generates more tortuous pathway for reactant transport and greatly reduces the effective diffusivity. Tortuosity of CL is much higher than conventional Bruggeman equation adopted. Knudsen diffusion plays a significant role in oxygen diffusion and significantly reduces the effective diffusivity. Reactive transport inside the CL is also investigated. Although the reactive surface area of non-precious metal catalyst (NPMC) CL is much higher than that of Pt CL, the oxygen reaction rate is quite lower in NPMC CL compared with that in Pt CL, due to much lower reaction rate. Micropores (a few nanometers) in NPMC CL although can increase reactive sites, contribute little to enhance the mass transport. Mesopores (few tens of nanometers) or macropores are required to increase the mass transport rate.

# 1. Introduction

Hydrogen has been considered as a promising future energy that can substitute for conventional fossil fuels. As a hydrogen utilizing device, the proton exchange membrane fuel cell (PEMFC) has drawn much attention due to its unique advantages and potential for portable, vehicular and stationary applications. However, several technical barriers still needs to be conquered before widespread commercialization of PEMFC, such as high cost, material durability and water and heat management. The utilization of precious metal catalyst, platinum Pt, which is for accelerating the sluggish electrochemical reactions, contributes to a major proportion of the cost of PEMFC. Most Pt is loaded at the cathode side, due to the inherently sluggish oxygen reduction reaction (ORR) in the cathode catalyst layer (CCL), whose reaction kinetics is five orders of magnitude slower compared to the hydrogen oxidation reaction (HOR) in the anode cathode catalyst layer [1]. Cost reduction can be realized by optimizing CCL design to enhance Pt utilization and thus to reduce Pt loading [2]. However, the limited resource and increased demand of Pt will steadily increase the price of Pt [3]. An alternative way is to develop well-performing non-precious metal catalysts (NPMCs) based on earth-abundant materials for ORR to replacing Pt-based catalysts, which can reduce the economic burden of fuel cell and further to gain commercial successes of PEMFC [2, 4-11].

Multiple physiochemical and electrical processes simultaneously occur in CCL, including reactant (oxygen or air) diffusion, electron and proton transfer, product (water) transport, and electrochemical reactions between oxygen, electron and proton, which are closely coupled and strongly interacted with each other. Consequently, for efficient functioning, at least four ingredients are indispensable in a CCL, the pore phase for reactant and product transport, the solid phase for electron conduct, the ionomer phase for proton transfer, and most important, the reactive sites for ORR reaction [12]. Conventional state-of-the-art CCL with precious catalysts are nanoscale porous media with random three-phase composites of pores, carbon-supported Pt (C/Pt) and ionomer (electrolyte) which provide pathways for oxygen, electron and proton transfer, respectively, where Pt nanoparticles with size of 2-5nm are finely dispersed on the carbon black with high surface area, serving as the reactive sites for ORR reaction [13]. The ORR reaction takes place at Pt particles surrounding by void, solid and ionomer phases, the so called triple-phase-boundary (TPB). For the NPMC, the scenario is a little complicated. Different approaches have been proposed to develop different kinds of NPMC materials for ORR. To date, the most promising NPMCs have been a class of pyrolysed transition metal nitrogen-carbon (M-N-C) complexes [4]. It is found that this kind of NPMC can be obtained after a heat-treatment of precursors including elemental transition ions (iron Fe and Cobalt Co in particular), carbon (carbon support, molecule, polymer) and nitrogen (MeN4 macrocycle, N-bearing molecule or polymer, N-containing gas) [4, 5]. The exact nature of the active ORR reactive sites in such M-N-C catalysts is still under debate, especially for the role of transition metal [10, 14]. The nitrogen species doped in the carbon structures are likely critical to the active site performance. Unlike precious metal catalyst layer, there is no distinction between catalyst and the support in M-N-C NPMC, in other words, the catalyst and the support are the same. Advanced experimental probes and atomistic simulations are urgently required to identify the active sites at the atomic level.

To replace precious catalysts, the activity, performance and durability of the NPMCs should be improved [4]. The volumetric activity of an NPMC is the product of the active site density and the single-site activity [15]. With extensive efforts, NPMC volumetric activity has undergone significant improvements by deliberately selecting the synthesis and treatment techniques, especially during the past few years [7, 8]. Note that the volumetric activity is measured under relatively high potential (0.8V). Under such circumstance, the only limited factor is electro-catalytic process. However, when fuel cell operates at low potentials or high current densities, transport limitations play roles. Thus, NPMC activity must be distinguished from its performance [4]. To achieve high activity, the thickness of NPMC CL is usually ten times thicker than that of Pt CL, leading to high mass transport resistance [5]. The maximum power density of fuel cells using NPMC is still required to be improved to be capable of the current Pt CL [4, 5, 9, 10]. To data, the researches of NPMCs have been focusing on enhancing the NPMC activity and there are quite few studies particularly dealing with transport issues in NPMCs [9]. For practical employment of NPMCs in fuel cells, therefore, enhancing the transport efficiency within the NPMC CL is urgently required to prompt successful adoption of NPMC and commercialization of PEMFC.

The successful development of CCL significantly depends on the fundamental understanding of the complex, multicomponent nature of the CCL as well as the physicochemical reactions in the CCL. Given the nanoscale characteristics, it is extremely difficult to implement in situ investigations of reactive transport processes experimentally. Therefore, numerical simulations become the most promising method for gaining meaningful insights of the complex, interconnected electrochemical processes occurring in CCL. Besides, numerical studies is less time consuming, inexpensive and especially efficient when systematic investigations involving several parameters are implemented. Over the past two decades, numerical investigations of the Pt-based CCL moved from ultra-thin interface models [16], to homogeneous models [17], to agglomerate models [18], and most recently to pore-scale models [19-27]. Each step advances with a higher resolution of the nanoscale porous structures and a requirement for more complicated numerical models and methods [18], helping in gaining deeper insights about the processes influencing the cell performance. Particularly, pore-scale simulations are based on explicitly resolving the realistic microstructures of porous media and describe the transport processes without making assumption about the microstructures on the macro transport properties. They can be used to improve the fundamental understanding of the reactive transport processes in porous media. The growing evidence of the importance of pore-scale simulations of PEMFC is reflected in the increasing interest in pore-scale modeling fluid flow and transport phenomena [19-31]. Two essential steps are required in the pore-scale simulations, namely the accurate descriptions of the complex porous structures and the establishment of pore-scale simulation methods. Reconstruction of the CL structures is an efficient technique for the former step and several reconstruction schemes have been developed in the literature [21-23]. For the latter one, different numerical methods have been adopted such as finite type method [22, 23, 32] and the lattice Boltzmann method (LBM) [21]. Due to its excellent numerical stability and constitutive versatility, the LBM has developed into an alternative and promising numerical approach for simulating fluid flow and transport processes in recent years, and is particularly successful in applications involving interfacial dynamics and complex geometries, e.g. multiphase or multicomponent flows in porous media [33-39].

The objective of the present study is to develop a pore-scale model based on the LBM for the complicated coupled physic-electrochemical processes in the CCL of PEMFCs. To the best of our knowledge, only Kim et al. adopted LBM for transport processes in CL and they did not consider the coupled electrochemical reaction processes [21]. The purpose of the present study is to apply the pore-scale model to evaluate the effects of structural characteristics on the macroscopic transport properties and to improve the fundamental understanding the reactive transport processes in the nanostructures of CCL using Pt catalysts and NPMCs. Effects of the reactive sites distributions and micropores and mesopores in the NPMCs on the mass transport and cell performance are investigated in detail. To the best of our knowledge, this is the first work about pore-scale studies of NPMC CL are performed.

The remaining part of the present study is arranged as follows. In Section 2, the reconstruction processes of nanostructures of CL are introduced. In Section 3, the physico-electrochemical model for reactive transport processes within the CL is established. In Section 4, the numerical method LBM is briefly introduced. In Section 5, effects of the nanostructures of the CL on the effective transport coefficients of oxygen and proton are investigated. The predicted effective transport coefficients are compared with results in the literature. Besides, reactive transport processes in CL with Pt catalyst and NPMCs are studied and compared. Finally, some conclusions are drawn in Section 6.

## 2. Reconstruction of nanostructures of the catalyst layers

Before performing the numerical simulations, the nanoscale topology of the CL must be obtained, which is one of the most important steps towards accurate prediction of the nanoscale reactive transport processes in CL. This can be achieved by computer tomography using experimental techniques such as SEM and TEM [13, 40]. Alternatively, reconstruction algorithm provides an efficient way to generate the nano-scale structures of CL. Several reconstruction methods have been developed in the literature. The first one is the Gaussian random field method proposed by Wang et al. [19, 20]. In this method, a random number generator was adopted to assign phases to voxels in a discrete and digitized domain representing CL. However, the nanoscale structures of the CL was resolved in a very coarse level, with only two phases existing including the void phase and another phase with mixed ionomer and C/Pt. Similar method was also adopted by Zhang et al. [32]. Slightly different from [19, 20], in [32] one phase consists of mixed ionomer-pore phase and the other phase is C/Pt. Employing such reconstruction method, transports of oxygen, electron or proton were modeled using an effective transport coefficient in the mixed phase based on their volume fractions [19, 20, 32]. The second method is a sphere-based annealing method developed by Kim and Pitsch [21], in which the reconstruction problem is formulated as the optimization problem to find a objective structure that satisfies some specified statistical characteristics of a reference structure. In [21], the CL is represented by carbon spheres which is allowed to move during the reconstruction process to achieve a prescribed porosity and a two-point correlation function. The carbon sphere is coved by thin ionomer films whose thickness can be determined by the Nafion content. The third method can be called random carbon sphere method in which carbon spheres are randomly distributed in the domain with specified overlap probability and overlap tolerance [22, 25, 26]. Similar to [21], a thin Nafion film of a given thickness is covered on the surface of the carbon spheres [22, 25, 26]. Lastly, Siddique and Liu [23] reported a process based reconstruction method in which the nanoscale structures of the CL is reconstructed step by step based on the experimental fabrication process including carbon seed generation, carbon phase growth, Pt deposition and Nafion coverage. With high resolution of 2 μm, detailed distribution of Pt particles are taken into consideration, which is ignored in previous reconstruction work [19-22, 25, 26, 32].

2.1 Precious metal catalyst layer

For fully resolving the nanostructures of Pt CL with detailed distributions of carbon black, Pt particles and Nafion, the Gaussian random field method is not possible as certain two phases are completely mixed without identifying the phase interface in the reconstructed structures [19, 20, 32]. The sphere-based annealing method, the random carbon sphere method and the process based reconstruction method can be adopted for this purpose. The sphere-based annealing method usually takes a long time for converging to the target structures [21]. In the random carbon sphere method, each carbon sphere is individually distributed in the domain and agglomerates are not likely to be formed , although the spheres are allowed to overlap to some extent [22, 25, 26]. SEM images of CL clearly indicate formation of agglomerates with typical size of several hundreds of nanometers [23]. Therefore, in this work, we follow the idea of process based reconstruction method proposed by Siddique and Liu [23] and adopt a reconstruction method called quartet structure generation set (QSGS) [41] for generating carbon and Nafion distributions. Besides, Pt particles are considered and a random number generator is employed to randomly disperse Pt particles on the surface of solid carbon. Before the reconstruction processes, the volume fractions of different phases are required and are determined in the following section.

*2.1.1 Volume fractions of different components*

CL manufacturing are usually controlled by the following design parameters including: (1) the platinum load ($\gamma_{Pt}$, mg cm$^{-2}$); (2) ionomer load ($\beta_N$, weight ratio of Nafion to total electrode weight), (3) platinum to carbon weight ratio ($\theta_{Pt}$) [24]. With these information, volume fractions for different phases, which are required for determine the number of cells occupied by each phase in the digitized reconstructed structures, can be calculated. Suppose the total weight of a CL is $m$, the volume of each phase is

$$\frac{m\beta_N}{\rho_N} = V_N \tag{1a}$$

$$\frac{m(1-\beta_N)(\frac{\theta_{Pt}}{1+\theta_{Pt}})}{\rho_{Pt}} = V_{Pt} \tag{1b}$$

$$\frac{m(1-\beta_N)(\frac{1}{1+\theta_{Pt}})}{\rho_C} = V_C \tag{1c}$$

where subscripts "Pt", "C" and "N" represent platinum, carbon and Nafion, respectively. Determination of the volume fractions of different phases in CL requires a designed value of the porosity $\varepsilon$, or the void space volume fraction, which accounts for the void pore space available for the mass transport. The porosity must remain above a certain value to guarantee that the pores in the CL are connected and thus the reactant can percolate through the entire CL. Obviously, the sum of the volume of Pt, Carbon and Nafion is equal to the total volume not occupied by void space

$$V_{Pt} + V_C + V_N = V(1-\varepsilon) \tag{2}$$

where $V$ is the total volume of the domain. Substituting Eq. (1a-1b) into Eq. (2) with $m$ expressed by $V_N$, the following equation can be obtained

$$\frac{\frac{V_N \rho_N}{\beta_N}(1-\beta_N)(\frac{\theta_{Pt}}{1+\theta_{Pt}})}{\rho_{Pt}} + \frac{\frac{V_N \rho_N}{\beta_N}(1-\beta_N)(\frac{1}{1+\theta_{Pt}})}{\rho_C} + V_N = V(1-\varepsilon) \quad (3)$$

Thus, the volume fraction of Nafion can be calculated based on $\beta_N$, $\theta_{Pt}$ and $\varepsilon$

$$\varepsilon_N = \frac{V_N}{V} = (1-\varepsilon)\frac{(1+\theta_{Pt})\beta_N/(\rho_N(1-\beta_N))}{\theta_{Pt}/\rho_{Pt}+1/\rho_C+(1+\theta_{Pt})\beta_N/(\rho_N(1-\beta_N))} \quad (4a)$$

The volume fractions of Pt and carbon can be obtained in the similar way as

$$\varepsilon_C = \frac{V_C}{V} = (1-\varepsilon)\frac{1/\rho_C}{\theta_{Pt}/\rho_{Pt}+1/\rho_C+(1+\theta_{Pt})\beta_N/(\rho_N(1-\beta_N))} \quad (4b)$$

$$\varepsilon_{Pt} = \frac{V_{Pt}}{V} = (1-\varepsilon)\frac{\theta_{Pt}/\rho_{Pt}}{\theta_{Pt}/\rho_{Pt}+1/\rho_C+(1+\theta_{Pt})\beta_N/(\rho_N(1-\beta_N))} \quad (4c)$$

Finally, the platinum load $\gamma_{Pt}$ (mg cm$^{-2}$) is related to $\varepsilon_{Pt}$ by the following equation

$$\gamma_{Pt} = \varepsilon_{pt}\rho_{Pt}l \quad (5)$$

where $l$ is the thickness of the CL.

### 2.1.2 Reconstruction processes

#### 2.1.2.1. QSGS for generating carbon

The QSGS is employed for carbon seed generation and carbon growth. The QSGS has been demonstrated to be capable of generating morphological features closely resembling the forming processes of many real porous media [41]. The procedures are briefly described as follows

(1) Randomly locate the seeds of carbon in the domain based on a distribution probability, $c_d$, which is smaller than the volume fraction of carbon in the rock, $\varepsilon_C$. A random number within (0,1) will be assigned to each cell in the domain, and if the random number is no greater than $c_d$ the node is selected as a carbon core.

(2) Grow an existing carbon cell (for example, $A$ with position $(i, j, k)$ ) to its neighboring cells,. For a neighboring cell in $\alpha$ direction, $A_\alpha$:

a. if one of its neighboring cells except cell $A$ is an carbon phase cell, $A_\alpha$ is accepted as a carbon cell, since materials tend to bond with each others to reduce the surface energy;

b. Otherwise, $A_\alpha$ is accepted based on another probability $P_\alpha$. In the present study, only the nearest neighboring cells are considered, namely six in the main direction (1,2,3,4,5,6), as shown in Fig. 1. For each direction, a new random number is generated, and if it is no greater than $P_\alpha$ the neighboring cell in $\alpha$ direction becomes a carbon phase cell.

(3) Repeat (1) and (2) until the volume fraction of carbon phase is achieved.

There are three parameters ($c_d$, $P_\alpha$, $\varepsilon_C$) controlling the generated morphologies of the carbon phase. $c_d$ denotes the number density of the growing seeds of carbon in the CL. Higher value of $c_d$ leads to more seeds in the system, resulting in more particles of carbon phase, while lower value leads to less seeds, generates bigger carbon agglomerates and thus would increase the computation fluctuation [41]. $P_\alpha$ is the probability of the growth of an existing carbon cell towards its neighboring cell in $\alpha$ direction. In order to generate isotropic structures, $P_\alpha$ of the six main directions (1,2,3,4,5,6) should be the same. Antistrophic structures can be obtained by simply setting different values of $P_i$ in different directions.

*2.1.2.2. Random dispersion of Pt particles*

With the help of advanced techniques such as TEM, it is revealed that the Pt particles are randomly distributed throughout the solid phase of the CL [13]. The Pt nanoparticles usually have a typical size of 2-5 nm [42]. Based on the above experimental observations, Pt particles with size of 5 nm, which is the resolution of a computational cell in the present nanoscale simulations, are randomly located on the surface of the carbon phase generated in Section 2.1.2.1. A random number generator, which can generate random numbers in the interval (0,1) with equal probability, is applied to each cell adjacent to the carbon phase in the domain. If the number assigned to a cell is not larger than a prescribed probability, which is chosen as the volume fraction of Pt in this work, this cell is determined as a Pt phase cell. This process is repeated until the volume fraction of Pt is achieved. With the above distribution scheme, generally, five faces of a cubic Pt nanoparticle are exposed to void or ionomer phase, while one face is contacted with the carbon support.

*2.1.2.3. QSGS for generating Nafion*

In the simplest case, the Nafion can be assumed as a thin film uniformly covering on the surface of C/Pt. Most of the agglomerate models [18] and quite a few pore-scale models [21, 22, 25] are based on this scenario. Lange et al. [22, 25, 26] pointed out that using a uniform distribution of thin Nafion film may lead to overestimated effective oxygen diffusivity and underestimated proton conductivity [25]. When nonuniform Nafion is presented, much narrow void space can be formed which acts as barriers for reactant transport. In addition, Nafion with nonuniform distributions also tends to connect between carbon spheres which can reduce the turtuosity, thus resulting in enhanced proton conductivity [25]. In this work, QSGS is adopted to generate the non-uniform distributions of Nafion on the surface of C/Pt. The generating steps are similar to that in Section 2.1.2.1 for carbon phase, except that the generation of Nafion seed is not requited as the existing C/Pt cells can be considered as the seeds. Note that the scheme introduced in Step 2a in Section 2.1.2.1 helps to establish connections between different local parts of Nafion, which is expected to increase the effective conductivity of Proton.

2.1.3. Characterization of the CL nanostructures

Characterization of the CL nanostructures includes porosity, pore size distribution (PSD), specific surface area and connectivity of different phases. These characteristics are closely related to the transport properties and reactive processes in the CL. Understanding these relationships can help to optimize the CL nanostructures and thus to improve the cell performance.

In the present work, all the reconstructed CL structures are based on the following parameters: $\beta_N$=0.3, $\theta_{Pt}$=0.4, $\rho_N$=2000 mg cm$^{-3}$, $\rho_C$= 1800 mg cm$^{-3}$ and $\rho_{Pt}$= 21450 mg cm$^{-3}$ [24]. We reconstruct different CL structures by changing the porosity. Note that changing the porosity affects several other parameters including volume fractions of C, Pt and Nafion, as well as the Pt loading; and larger porosity leads to smaller volume fractions of C, Pt and Nafion and lower Pt loading.

Fig. 2(a) shows a 3D structures of CL reconstructed using the above method with $\varepsilon$=0.4. The size of the domain is 4000×500×500 nm, discritezed by 800×100×100 cells with resolution of 5 nm per cell. For reconstructing this structure, only about ten minutes are required on a Dell T7400 Workstation with eight processors (Intel Xeon E5420) and 3GB of memory. The volume fractions of carbon, Pt and Naftion calculated using Eq. (4) is 38.12%, 1.28% and 20.59%, respectively. The ionomer-to-carbon weight ratio (I/C ratio) is 0.60. The corresponding Pt load calculated using Eq. (5) with typical CL thickness of 4 μm is 0.11 mg cm$^{-2}$. As mentioned above, the size of a Pt nanopaticle is 5nm, indicating a Pt particle can completely occupy a computational cell. In the reconstructed structures shown in Fig. 2, there are totally 104443 Pt nanoparticles. Fig. 2(b) shows the detailed distributions of carbon black (top), C/Pt (middle) and Nafion covered on C/Pt (bottom) in 200<$x$<500 (lattice units) at the center of $z$ axis. The nonuniform coverage of Nafion on the surface of C/Pt can be clearly observed.

PSD is an important structure characteristic of porous media. Determining the pore size of each pore cell is also required for the calculating of the Knudsen diffusivity which will be discussed in Section 3. The 13 direction averaging method proposed in Ref. [22] is adopted to calculate pore size of each pore cell. Here 13 directions are 3 directions along the $x$, $y$ and $z$ axis, 6 diagonal directions in $x$-$y$, $x$-$z$ and $y$-$z$ planes with 2 diagonal directions in each plane, and 4 diagonal directions traversing the 3D space. For each pore cell in a certain direction, a pore length, which is defined as the length of a line occupied by pore cells only, is determined. The pore size of the current pore cell is calculated as the averaged value of the pore lengths in the 13 directions. Fig. 3 shows the PSD of the reconstructed CL structures shown in Fig. 2.

Experimental measurements of PSD of CL have indicated that the PSD of CL is bimodal. Using systematic nitrogen adsorption method, it was found that two local peaks of the relative volume can be found, with one below 5 nm and another at around 50~60 nm [43]. Since the resolution of one cell in the present work is 5nm, it is impossible to capture the peak below 5 nm, leading to unimodal distributions as shown in Fig. 3. The maximum of the PSD is located at around 50 nm, in good agreement with the experimental results [43]. All the pore sizes are located in the range of 5-130 nm, indicating the nanoscale characteristics of the CL.

The total surface area the total volume of the carbon phase are 2782840 $(\delta x)^2$ and is 3230161$(\delta x)^3$, respectively. The determination of the specific surface area obviously depends on the resolution of each cell $\delta x$, and lower $\delta x$ leads to higher specific surface area. With $\delta x$=5nm the specific surface area is 86.2 $m^2\ g^{-1}$ for the reconstructed CL shown in Fig. 2. This value is relatively lower than the experimental results in [43], where the specific surface area for Ketjen Black carbon and Vulcan XC-72 are 890±64.3 $m^2\ g^{-1}$ and 228±3.57 $m^2 g^{-1}$. Failing capture of pores with size smaller than 5nm is a possible reason for underestimated specific surface area in the present study. As reported in Ref. [43], pores of size <2nm constitute ~25% of the total pore volumes of Ketjen Black carbon, which obviously contribute a lot to the specific surface area.

Connectivity of each phase strongly affects the effective active sites. In the CL, electrochemical reactions only take place at Pt particles around which computational cells of void phase, C/Pt and ionomer are "transport". Here, a cell is "transport" means that this cell belongs to a continuous percolation path throughout the entire CL. It is obvious that if a portion of a certain phase does not penetrate the entire CL, the corresponding phase entering the CL from one end (For reactants, proton and electron, they enter the CL from GDL/CL interface, CL/PEM interface and GDL/CL interface, respectively) of the CL cannot arrive at the outlet through this portion which thus is "dead" [20]. For identifying the "transport" and "dead" portions in the CL, the connectivity of a certain phase is required to be determined. Followed previous work in [44], a connected phase labeling algorithm is developed to identify the "transport" and "dead" portions and to count the number of cells in each portion. This algorithm scans the entire domain, checks the connectivity of a cell with its neighboring 26 points (as shown in Fig. 4a), labels the cell depending on the local connectivity, and finally assigns any portion with a distinct label if it is disconnected from other portions in the domain. The general steps for labeling cells belonging to a certain phase, for example solid phase C/Pt, in a domain with minimum location (1,1,1) and maximum location (*nx,ny,nz*) are as follows

(1) Scanning the domain from (1,1,1) to (nx,ny,nz) by moving along *x*, *y*, *z* directions. For the current node *P* under inspection, checking the 13 neighboring cells (blue cells in Fig. 4(a)) which have already been scanned:

(a) If none of the 13 neighboring cells is solid phase, assign a new label to P;

(b) If one or more than one cells belong to solid phase, find the smallest labels of these solid phase cells and assign the smallest label to P and these solid phase cells.

At the end of this step, each solid phase cell in the system has been labeled. However, it is very likely that connected cells may be assigned with different labels. Thus Steps 2-3 are further implemented.

(2) Scanning the domain from (nx,ny,nz) to (1,1,1) by moving along -*x*, -*y*, -*z* directions. For a current solid phase cell P, searching solid cells in its 26 neighboring cells, determining the smallest label of and assign this value to P and its neighboring solid phase cells.

(3) Scanning the domain from other six diagonal directions, and using the same scheme in Step 2 to update the label.

Steps 2-3 can guarantee that a distinct label is assigned to any portion of solid phase if it is disconnected from any other portions of solid phase in the domain. The portion which have cells located at *x*=1 and *x*=nx, which indicates that this portion penetrates the entire domain, is called a continuous percolation path throughout the CL, and cells in this portion is "transport". Otherwise, a portion does not penetrate the entire domain and cells belonging to this portion are "dead". Through such connected phase labeling algorithm the "transport" cells for a certain phase can be determined and the surface area and volume of each portion can be calculated. Fig. 4b shows the C/Pt phase after labeled. The black portion represents the "transport" C/Pt while the red portions inside the domain denote the "dead" portions. The results show that there are 443 portions of C/Pt in the reconstructed CL with one of them as "transport" and the remaining as "dead". The "transport" portion occupies most of the C/Pt phase, with a volume fraction of

99.80% (3328017/3334604, 3334604 is the total number of C/Pt cells), as clearly seen in Fig. 4b where the "dead" portions present as small and discrete blobs. For the void phase, there are totally 536 portions in the domain with one "transport" portion and 535 "dead" portions, and the volume fraction of the "transport" portion is 99.69% (2892039/2901053). For the ionomer, 793 portions exist in the domain with one "transport" portion and 792 "dead" portions, and the volume fraction of the "transport" portion is 99.86% (1761838/1764343). Connectivity analysis of reconstructed CL based on SEM in a recent experiment revealed 99.8% of solid phase and 99.9% of the void phase are "transport" [13]. The results of the present study are in good agreement with the experimental results. The connectivity analysis of ionomer in Ref. [13] was not reported as SEM technique cannot distinguish the solid phase and ionomer. High "transport" characteristics of different phases in the present study benefit from the Step 2a of QSGC in Section 2.1.2.1.

2.2. Non-Precious metal catalyst layer

Today, M-N-C catalysts obtained from the pyrolysis of precursors including transitional metal, nitrogen and carbon are the most promising NPMC for the ORR [4, 7]. A large degree of flexibility exists for developing M-N-C catalysts using different nitrogen, transition metal and carbon sources. The final nanostructures and performance of M-N-C catalysts are strongly affected by several factors including the types of precursors, heat treatment temperature, carbon support morphology and synthesis conditions [10]. Therefore, we need to select a NPMC from the varied NPMCs in the literature. The NPMC developed by Wu et al. [10] is studied, which is a breakthrough of NPMC developments. The so called PANI-FeCo-C is developed as follows as shown in Fig. 5. First, Ketjenblack, aniline oligomer and transition metal salts were mixed. Ammonium then was added into the mixture to polymerize aniline. Then, the composite was dried and subjected to the first time heat treatment in nitrogen in the range 400-1000 $^o$C. The heat treated product was preleached using acid, after which a second time heat treatment in nitrogen was implemented. The NPMC based on Polyaniline and mixed-FeCo shows excellent performance, achieving a maximum power density of 0.55W cm$^{-2}$ at a cell voltage of 0.38V, and excellent performance retention after 700 h of potential hold at 0.4 V [10]. Fig. 6 shows a SEM of the PANI-FeCo-C with heat treatment of 900°C [10]. Comparing the SEM with that of Pt CL shown in Ref. [23] (Fig. 1 in Ref. [23]), it can be found that the porous structures of the two different catalysts at the scale of hundreds nanometers are quite similar, showing formation of agglomerates with approximately equal size of primary particles. This observation leads to a strong yet reasonable assumption for studying NPMC CL in the present work. We assume that the NPMC CL studied has the same porous structures with Pt CL, and we focus on the effects of the distributions of reactive sites.

One may argue that the ingredients of the solid phase of PANI-FeCo-C CL are quite different from Pt CL. For the PANI-FeCo-C CL, the solid phase consists of carbon black, graphite PANI and metal aggregates [10], much complicated than that of Pt-based CL. However, since the electron conductivity is much higher than proton conductivity within the ionomer, the potential loss within the solid phase can be neglected, compared with that in ionomer, which is a commonly adopted assumption for studying reactive transport in CLs [19, 23]. Therefore, the detailed ingredients of the solid phase can be out of consideration.

Using PANI as the source of nitrogen and carbon has several benefits. The similarity between the structures of PANI and graphite facilitates the incorporation of nitrogen-containing active sites into the partially graphitized carbon matrix during the heat treatment. Besides, using PANI can guarantee the uniform distributions of nitrogen sites on the surface of solid phase and thus an increase in active-site density [10]. Hence, in this work for the PANI-FeCo-C CL, the active sites uniformly distribute on the entire surface area of solid phase, but with a much lower reaction rate compared with that of Pt [10]. In short, the C/Pt solid phase generated in Section 2.1 is considered as a mixture of carbon black, graphite PANI and metal aggregates in PANI-FeCo-C CL, and the entire solid surface is active sites. The catalyst loading, based on an assumed solid density of 1800 mg cm$^{-3}$, is 0.283 mg cm$^{-2}$.

## 3. Model description

There have been several pore-scale simulations of CL concerning different aspects of electrochemical processes [19-27]. Structural resolution in [19, 20, 27] is quite low due to mixture phase adopted. In [21, 22, 25, 26] because the pore-scale phenomena on the order of 100 nm are of primary interest, the detailed distributions of platinum particles, which is with typical size of 2~5 nm, were neglected. To the best of our knowledge, only in [23] detailed distributions of Pt nanoparticles were taken into consideration and the Pt utilization efficiency were investigated under different structural parameters. In the present study, for Pt CL, the detailed distribution of Pt nanoparticles is resolved. For NPMC-based CL, the entire surface of solid phase is reactive.

The computational domain of the pore-scale simulations is CL reconstructed based on the method in Section 2. The spatial coordinate $x$ is defined so that the positive direction points from the gas diffusion layer (GDL)/CL interface to the CL/proton exchange membrane (PEM) interface, with its origin located at the GDL /CL interface. The interconnected physico-electrochemical processes modeled are as follows: the oxygen diffusion in the GDL and the CCL; the oxygen dissolution at the pore-ionomer interface; the diffusion of dissolved oxygen in the ionomer covered on the solid (catalyst/C) phase; the proton conduction in the ionomer; the electrochemical reactions of ORR on the reactive sites.

The key assumptions employed in this model are as follow.a) Electron conduction within the solid phase is not considered, as the electronic conductivity is sufficiently high and the potential in the solid can be assumed to be uniform [19] ; b) Effects of water vapor and liquid water flow and distributions are not considered [19, 22]; c) Isothermal and steady state operation; d) Constant conductivity of proton in the ionomer. Details of the model are as follows, for more information one can refer to [19, 22, 23, 25, 26].

3.1 Oxygen diffusion in the pores

Oxygen diffusion in the pores of CL is mainly due to the oxygen concentration gradient and is approximately modeled by Fick's law with the following equation:

$$\nabla \cdot (D_{O_2} \nabla C_{O_2}) = 0 \tag{6}$$

where $C$ is the concentration and $D$ is the diffusivity. Binary diffusion for oxygen into nitrogen is assumed, and the bulk diffusivity is calculated by [45]

$$D_{O_2,b}(\text{m}^2\ \text{s}^{-1}) = 0.22 \times 10^{-4} \frac{(T/293.2\text{K})^{1.5}}{p/1\text{atm}} \tag{7}$$

where $p$ it the pressure of the CL. When the character length of a system is comparable to or smaller than the mean free path of the molecules involved, collisions between molecules and solid wall are more frequent than that between molecules, and such mean of diffusion is called Knudsen diffusion. The Knudsen diffusivity of oxygen is calculated by the following equation according to the dusty gas model [46]

$$D_{O_2,Kn}(\text{m}^2\ \text{s}^{-1}) = \frac{d_p(\text{m})}{3}\sqrt{\frac{8RT}{\pi M_{O_2}(\text{Kg mol}^{-1})}} = 48.5 d_p(\text{m})\sqrt{\frac{T}{32}} \tag{8}$$

where $R$ is the gas constant and $d_p$ is the local pore diameter for a pore cell in the domain. In the present study, $d_p$ is an effective pore diameter calculated by the 13 direction averaging method proposed in Ref. [22]. All the variables in Eq. (8) are of SI units. The total oxygen diffusivity is computed using Bosanquet formula by [47]

$$D_{O_2}(\text{m}^2\ \text{s}^{-1}) = (\frac{1}{D_{O_2,Kn}} + \frac{1}{D_{O_2,b}})^{-1} \tag{9}$$

3.2 In the ionomer

Oxygen is transported to the reactive sites as a molecular species dissolved in the ionomer. The dissolved oxygen concentration at the pore-ionomer interface, $C_{O_2,e}$, is determined by Henry's law

$$C_{O_2,e} = \frac{C_{O_2}RT}{H} \tag{10}$$

$H$ is the Henry's constant and is calculated by [48]

$$H(\text{Pa m}^3\ \text{mol}^{-1}) = 1.01325 \times 10^{-1} \exp(-666/T + 14.1) \tag{11}$$

The diffusion of the dissolved oxygen in the ionomer is governed by the following equation

$$\nabla \cdot (D_{O_2,e} \nabla C_{O_2,e}) = S_{O_2,e} \qquad (12)$$

where $S$ is the source term related to the ORR reaction and will be described later. $D_{O_2,e}$ is the oxygen diffusivity in the electrolyte and is calculated by [22]

$$D_{O_2,e}(m^2\ s^{-1}) = 10^{-10} \times (0.1543(T-273) - 1.65) \qquad (13)$$

As the concentration field is discontinuous at the pore-ionomer interface, one way is to solve Eq. (6) and Eq. (12) separately in each domain with imposing Eq. (10) at the interface. Another efficient way is to transfer Eq. (12) using Eq. (10). Instead of solving for the real concentration in the ionomer phase, a hypothetical concentration $C_{O_2}$ in this phase is solved. Based on the isothermal condition of this work, combining Eq. (10) and Eq. (12) yields the following equation

$$\nabla \cdot (D_{O_2,e} \nabla C_{O_2}) = \frac{H}{RT} S_{O_2,e} \qquad (14)$$

Therefore, a continuous filed $C_{O_2}$ can be defined in the pore and ionomer, with its governing equation as the following form

$$\nabla \cdot (D \nabla C_{O_2}) = S_{O_2} \qquad (15)$$

with $D$ as $D_{O_2}$ and $D_{O_2,e}$, and $S$ as 0 and $S_{O_2,e} H/RT$ in pore and ionomer phase, respectively. After the simulation, the actual concentration in the ionomer can be back-calculated using Eq. (10).

3.3 Proton conduction in the electrolyte

The governing equation for the proton transfer in the electrolyte is as follows

$$\nabla \cdot (\sigma \nabla \varphi_e) = S_e \qquad (16)$$

where $\sigma$ is the proton conductivity and $\varphi$ is the potential of the electrolyte phase.

3.4 ORR reaction rate

The ORR rate, $i$, is a function of local oxygen concentration and potential. It is described by the Tafel equation as follows

$$i = i_0 \frac{C_{O_2,e}}{C_{O_2,ref}} \exp(-\frac{\alpha F}{RT}\eta) \tag{17a}$$

where $C_{O_2,ref}$ is the reference oxygen concentration, $\alpha$ is the cathode transfer coefficient, $F$ is the Faraday's constant and $i_0$ is the exchange current density. As $\eta = \varphi_s - E_{eq} - \varphi_e$, and solid phase potential $\varphi_s$ and equilibrium potential $E_{eq}$ are assumed to be constant in the present work [20], the Tafel equation can be modified as

$$i = i_{0,m} \frac{C_{O_2,e}}{C_{O_2,ref}} \exp(\frac{\alpha F}{RT}\varphi_e) \tag{17b}$$

$i_{0,m}$ is the modified exchange current density. Therefore, the source term of Eq. (15) and Eq. (16) is [23]

$$S_{O_2} = \begin{cases} \frac{H}{RT} \frac{n}{\Delta} \frac{i}{4F} & \text{in an active active ionomer cell} \\ 0 & \text{others} \end{cases} \tag{18a}$$

$$S_e = \begin{cases} \frac{n}{\Delta} i & \text{in an active active ionomer cell} \\ 0 & \text{others} \end{cases} \tag{18b}$$

where $n$ is the total number of neighboring cells occupied by Pt nanoparticles and $\Delta$ is the length of a computational cell.

3.5 Boundary conditions

3.5.1. Effective transport properties

For the prediction of oxygen effective diffusivity, constant concentrations are specified at the left and right boundaries. For the other two directions, namely $y$ and $z$ directions, periodic boundary conditions are employed. The effective diffusivity $D_{\text{eff}}$ along the $x$ direction is calculated by

$$D_{\text{eff}} = \frac{(\int_0^{L_y} \int_0^{L_z} (D\frac{\partial C}{\partial x})|_{L_x} \, dydz)/L_y L_z}{(C_{\text{in}} - C_{\text{out}})/L_x} \tag{19}$$

where $C_{\text{in}}$ and $C_{\text{out}}$ is the inlet and outlet concentration, respectively. $L_x$, $L_y$ and $L_z$ are the size of the domain in the $x$, $y$ and $z$ direction, respectively. $|_{L_x}$ means local value at $x=L_x$. For the effective conductivity of proton, the same boundary conditions can be applied and the effective conductivity of proton can be calculated using a similar equation to Eq. (19).

3.5.2. Coupled electrochemical reaction transport

We also perform simulations about the electrochemical reaction transport within the CL. The CL reconstructed here is considered to be adjacent to the GDL/CL interface. As reported in [32], for a 3μm thick CL simulated, the potential drop across the CL is just 0.02 V at a current density of 0.47 A cm$^{-2}$. In Ref. [22], a potential drop of 0.008 is considered across the porous CL. Our preliminary simulations by fully considering the oxygen transfer, proton conduction and electrochemical reactions, with the same boundary conditions adopted in [19, 32], shows that the overpotential drop across the 3μm thick CL shown in Fig. 2 is about 0.0118 V at a current density of 0.858 A cm$^{-2}$. Based on these results in the literature as well as our preliminary simulations, assuming a uniform distribution of ionomer potential in the entire CL is reasonable and practical, as fully resolving all the coupled processes in the complicated structures shown in Fig. 2 takes a long time to achieve convergence. In the present study, for the coupled electrochemical reaction transport, the transfer of proton in the ionomer is not solved and the potential in the entire CL is preset. Without loss of generality, the model described in Section 3.3 and 3.4 remains unchanged.

At the GDL/CL interface, one layer of pore-only cells is added to the computational domain as the reactant gas reservoir. The boundary conditions for oxygen are as follows

$$\text{At the right boundary:} \quad \frac{\partial C_{O_2}}{\partial x} = 0 \tag{20a}$$

$$\text{At the left boundary:} \quad C_{O_2} = C_{O_2,\text{GDL/CL}} \tag{20b}$$

For the other two directions, namely $y$ and $z$ directions, periodic boundary conditions are employed.

## 4. Numerical methods

Different numerical methods have been adopted to investigate electrochemical processes in CL of PEMFCs [19-27, 32]. To the best of our knowledge, only Kim and and Pitsch [21] and very recently Wu and Jiang [31] employed the LBM to investigate diffusion processes in CL; however, in both studies only transport processes were investigated, while electrochemical reactive processes were not taken into account [21]. In this work, for the first time the LBM is adopted to simulate the coupled processes of oxygen diffusion, proton conduction and electrochemical reaction taking place in CL of PEMFCs. Due to its excellent numerical stability and constitutive versatility, in recent years the LBM has developed into an alternative and promising numerical approach for simulating fluid flow and is particularly successful in applications involving interfacial dynamics and complex geometries [33, 34, 39]. LBM simulates pseudo-fluid particles on a mesoscopic level based on Boltzmann equation using a small number of velocities adapted to a regular grid in space. The obvious advantages of LBM are the simplicity of programming, the parallelism of the algorithm and the capability of incorporating complex microscopic interactions [33]. The LBM has been widely applied to the multiphase reactive transport in PEMFC [21, 27-29, 31, 39, 49].

For oxygen diffusion in void and ionomer phase, the evolution equation for the concentration distribution function is as follows

$$g_i(\mathbf{x}+\mathbf{c}_i\Delta t, t+\Delta t) - g_i(\mathbf{x},t) = -\frac{1}{\tau_g}(g_i(\mathbf{x},t) - g_i^{eq}(\mathbf{x},t)) + a_i S_{O_2} \Delta t \tag{21}$$

where $g_i$ is the distribution function with velocity $c_i$ at the lattice site **x** and time $t$. D3Q19 lattice model is adopted and the discrete lattice velocity $\mathbf{c}_i$ is given by

$$\mathbf{c}_i = \begin{cases} 0 & i=0 \\ (\pm 1,0,0), (0,\pm 1,0), (0,\pm 1,0) & i=1 \sim 6 \\ (\pm 1,\pm 1,0), (0,\pm 1,\pm 1), (\pm 1,0,\pm 1) & i=7-18 \end{cases} \tag{22}$$

The equilibrium distribution function $g^{eq}$ is defined

$$g_i^{eq} = C_{O_2}/a_i, \quad a_i = \begin{cases} 1/3 & i=0 \\ 1/18 & i=1 \sim 6 \\ 1/36 & i=7-18 \end{cases} \tag{23}$$

The concentration and the diffusivity are obtained by

$$C_{O_2} = \sum g_i, \quad D = \frac{1}{3}(\tau_g - 0.5)\frac{\Delta x^2}{\Delta t} \tag{24}$$

Eqs. (21) and (23) can be proved to recover Eq. (15) using Chapman–Enskog expansion.

For proton transfer in ionomer phase, a potential distribution function $h$ using D3Q19 lattice model can be defined, and the evolution equation, the equilibrium distribution functions and the relationship between collision time and proton conductivity are similar to that for oxygen concentration distribution functions. The related expressions are not repeated here for brevity [49].

## 5. Results and discussion

### 5.1 Effective diffusivity

Our code is validated by predicting the effective diffusivity of oxygen in a cubic domain containing a sphere whose diameter is the same as the side length of the cube [22]. 200*200*200 lattice is employed and the predicted effective diffusivity is 0.3317, in good agreement with the results in [22]. The details of the validation are not reported for brevity. As mentioned previously, for the first step study, no difference is designed between the nanostructures of the NPMC and Pt CLs. Thus, we discuss simulation results in the framework of Pt CL only, which is for the convenience to compare our predicted values with existing studies centering on Pt CL in the literature. For predicting effective transport properties, the entire domain is divided into 20, 4 and 4 subdomains in the $x$, $y$ and $z$ directions, respectively, leading to 320 CPU cores adopted for the code parallelized by MPI. The running time for proton conduction and oxygen diffusion without considering Knudsen diffusion is about 2 hours. When Knudsen diffusion is taken into account, the time is little longer of about 6 hours. The boundary conditions have been introduced in Section 3.5.1.

Fig. 7(a) shows the variations of effective diffusivity with porosity. In this section, all the reconstructed CL structures have the following characteristics: $\beta_N$=0.3, $\theta_{Pt}$=0.4, $\rho_N$=2000 mg cm$^{-3}$, $\rho_C$= 1800 mg cm$^{-3}$ and $\rho_{Pt}$= 21450 mg cm$^{-3}$ [24]. In Fig. 7, the effective diffusivity has been normalized by the bulk diffusivity $D_b$ given by Eq. (7). As the porosity increases, the volume fractions of solid phase and ionomer reduce. Therefore, it is expected that as the porosity rises, effective diffusivity of oxygen increases and effective conductivity of proton decreases, as shown in Fig. 7. The effective transport properties predicted by Lange et al. [22] are also plotted in Fig. 7. For the effective diffusivity, it can be observed that the values predicted in the present study are much lower than that of Lange et al. [22]. For example, for the effective diffusivity without considering Knudsen diffusion, at ε=0.4 the value predicted by the present study is 0.104, only about half that of Lang et al. [22] which is 0.195. The lower effective diffusivity in the present study, is exactly due to the nonuniform ionomer distributions generated by the structure reconstruction algorithm introduced in Section 2. As shown in the bottom image of Fig. 2(b), the thickness of ionomer covered on the surface of C/Pt is quite nonuniform, and the connections of ionomer between different parts of agglomerates even block the void space in some local regions. Such nonuniform distributions and connections greatly increase the tortursity of the path for oxygen transport in the void space, thus obviously reducing the effective diffusivity.

Macroscopic models of PEMFCs highly depend on empirical relationship between macroscopic transport properties and statistical structural information of porous components (permeability VS porosity, diffusivity VS porosity) including GDL and CL [17]. Bruggeman equation has been widely used in the macroscopic models to calculate the effective transport properties Γ, which has the following form

$$\Gamma_{eff} = \Gamma \frac{\varepsilon}{\tau} \qquad (28)$$

where ε is the volume fraction. $\tau$ is the tortuosity of a porous medium, and it is usually set as $\tau = \varepsilon^{-0.5}$, leading to effective transport properties as a pure function of the volume fraction, namely, $\Gamma_{eff} = \Gamma \varepsilon^{1.5}$ [17]. Following Eq. (28) and assuming a power relationship between ε and $\tau$ as $\tau = \varepsilon^{\alpha}$, the averaged $\alpha$ for the simulation results of the present work is -1.48, while that of Lang et al. [22] is -0.74 and that predicted very recently by Wu and Jiang is much lower as -2.11 [31]. Based on the above analysis as well as the effective diffusivity predicted by the Bruggeman plotted in Fig. 6, it can be concluded that the Bruggeman equation overestimates the effective diffusivity, and a lower $\alpha$ should be adopted in the Bruggeman equation to calculate effective diffusivity in CL of PEMFCs [22, 23, 31].

To the best of our knowledge, only Lang et al.[26], Kim and Pitsch [21] and Siddique and Liu et al. [23] considered the Knudsen diffusion process as well as the nonuniform distributions of the ionomer in the pore-scale study of CL. Fig. 7(a) displays the results from our simulations as well as that from the literature. Firstly, when Knudsen diffusion is further considered, the predicted effective diffusivity greatly declines, as shown in Fig. 7(a). This is expected based on Eq. (7-9). Under the simulation condition of $T$=333 K and $P$=1 atm, the binary diffusivity is $1.33 \times 10^{-5}$ m$^2$ s$^{-1}$. With a typical pore size of 60 nm, the local Knudsen diffusivity is about $9.39 \times 10^{-6}$ m$^2$ s$^{-1}$. Therefore, the total diffusivity in the local pore is $5.50 \times 10^{-6}$ m$^2$ s$^{-1}$ based on Eq. (9), much lower than the bulk diffusivity. In Ref. [26], Lang et al. compared their results with that of Kim and Pitsch [21] in detail. Both Lang et al [26] and Kim and Pitsch [21] adopted the sphere-based carbon structures, therefore the effects of structures on the transport properties have some similar ccharacteristics. Hence only the results of Lang et al. [26] (Fig. 4 in Ref. [26]) are displayed in Fig. 7(a). As shown in Fig. 7(a), both the present study and Siddique and Liu [23] predict lower values than that of Lang et al. [26]. Obviously, different reconstruction methods lead to quite different nanoscale structures of CL. As mentioned above, the present study and Siddique and Liu [23] adopted the fabrication process based reconstruction method, while Lang et al.[26] employed the random carbon sphere method. CL structures reconstructed using the former method (Fig. 2 in [23] and Fig.2 in the present study) is less regular than that using the later method (Fig. 1 in Ref. [22]), which is also proved by the tortuosity estimated in the previous paragraph, leading to lower effective diffusivity estimated. Therefore, it seems that the underlying structures of carbon black play an important role. Even the same reconstruction method is employed, the values of the effective diffusivity predicted by our simulations are quite smaller than that of Siddique and Liu [23], which are closer to the experimental results of General Motors Corporation [50]. The discrepancy is explained as follows. First, in step 2(a) of the QSGS, if one of the neighboring cells is an ionomer phase cell, the current cell is accepted as an ionomer cell, while in Ref. [23] at least two neighboring cells are required to be ionomer phase. Second, the ionomer load in the present study is 0.3 while that in Ref. [23] is 0.2. Third, a much thicker CL is simulated in the present study. The above three factors are believed to generate more tortuous pathways in the void space, result in lower effective diffusivity. The above results and discussions indicate that the detailed structures of CL, including the carbon structures as well as the distribution characteristics of ionomer, significantly affects the effective diffusivities in the CL [25]. Advanced structure reconstruction algorithm should be adopted to generate nanostructures more close to the realistic CL [23, 25]. Process based reconstruction algorithm is proved to be a good choice [23], based on the comparison between the simulation results of the present study and that in the literature. Further, it can be observed from Fig. 7(a) that as the porosity increases, the discrepancy between our simulation results and that from the literature reduces. For porosity in the range of 0.3~0.50, the effective diffusivity predicted by Siddique and Liu [23] ranges from 0.00927~0.0587, while that in the present study ranges from 0.000336 to 0.0239. The ratio at $\varepsilon=0.3$ is 27.59, while that at $\varepsilon=0.5$ is only 2.46. This is because Knudsen diffusion is very sensitive to the local pore size affected by the local ionomer distributions. Under low porosity, the local pore is quite likely to be blocked by the bridging of ionomer between two separated agglomerates. Finally, the simulation results in the present study, is still higher than the experimental results [50], and using more accurate relationship between pore diameter and Knudsen diffusivity is believed to further reduce the discrepancy between simulation and experimental results [26].

5.2. Effective conductivity of proton

Now attention is turned to the effective conductivity of proton in the ionomer. Fig. 7(b) shows the results of the present study as well as experimental and numerical results from the literature [25, 51]. The value of this work is very close to that with carbon sphere diameter $r$=20 nm and is lower than that with $r$=12 nm predicted by Lange et al. [25]. The reconstruction algorithm in the present study generates nonuniform distributions of ionomer. On one hand, the nonuniform distributions can improve the connection of ionomer between different solid phase agglomerates. Such connections can reduce the tortuosity of proton path and thus help to increase the effective conductivity. On the other hand, however, local extremely thin ionomer film on the surface of C/Pt severs as great barrier for proton transport, as shown in Fig. 2(c). Compared with the simulation results in Ref. [25], it can be seen that the nonuniform distributions of ionomer does not increase the proton conductivity. In addition, for solely considering proton transport in ionomer, again the Bruggeman equation overestimates the effective conductivity, just as its performance for predicting effective diffusivity as discussed previously, because it underestimates the tortuosity. Further, both this work and Ref. [25] as well as the Bruggeman equation predict a much lower value than the experimental results [51]. This indicates other mechanisms for proton transport in CL [25]. When PEMFC is operated under high current density, liquid water is generated in CL, which can increase the water content of ionomer and fills the void pores. Ionomer with high water content provides high proton conductivity. In addition, proton can directly transport in the liquid water [52]. Therefore, effective conductivity of proton not only depends on the ionomer content and distribution, but also depends on the operating conditions [51].

5.3 Coupled electrochemical reaction transport

The current balance for the cathode CL can be written as

$$\int_{\Gamma} i_0 \frac{C_{O_2,e}}{C_{O_2,ref}} \exp(-\frac{\alpha F}{RT}\eta) ds = i_d \cdot A_{cross} \quad (25)$$

where $A_{cross}$ is the cross-sectional area of CL and $i_d$ is the discharge current density applied on $A_{cross}$. The left hand side of Eq. (25) is the integration of the current density over all the reactive sites, with Γ denotes the interfacial surface over which the surface integral is taken. Assuming a averaged oxygen concentration and overpotenial over the entire CL, Eq. (25) can be modified as [19]

$$i_0 \frac{\overline{C_{O_2,e}}}{C_{O_2,ref}} \exp(-\frac{\alpha F}{RT}\overline{\eta}) \cdot \frac{A_{reaction}}{A_{cross}} = i_d \quad (26)$$

where $A_{reaction}$ is the total reactive surface are within the cathode CL. It can be concluded from Eq. (29) that for increasing $i_d$, one can increase the total reactive surface area $A_{reaction}$, raise the reaction rate (by increasing $i_0$), or enhance the mass transport inside the CL to rise $\overline{C_{O_2,e}}$. Compared with Pt-based CL where only the sites with Pt nanoparticle are reactive, for the PANI-FeCo-C CL studied, using PANI can guarantee the uniform distributions of nitrogen sites on the entire surface of solid phase [10], therefore the entire solid surface is considered active sites. Consequently, $A_{reaction}$ for PANI-FeCo-C CL is much higher than that for the Pt-based CL; and for the structure shown in Fig. 2, $A_{reaction, NPMC}/A_{reaction,Pt}$=7.44. However, the reaction rate for the non-precious metal CL is much lower with Pt [10], typically two orders of magnitude lower. In the present study, $i_{0,m}$ in Eq. (17) for PANI-FeCo-C is set as 1/100 of that of Pt [10]. Using the physicochemical model established in Section 3 and the LBM in section 4, reactive transport in Pt CL and PANI-FeCo-C CL is simulated. The parameters used in the simulations are listed in Table 1. The code developed is parallelized by MPI, and 4000 CPU cores are adopted in the simulation, which usually take 8 hours to converge.

The oxygen concentration field in the NPMC predicted by the LBM simulations is illustrated in Fig. 8, with $\varphi_e$ in the domain as 0.6V, which is a good value for studying both kinetic loss and mass transport limitations [4]. As shown in the figure, the distributions are rather complicated due to the complex nanoscale structures. The concentration continuously declines from the left inlet to the right outlet, due to the electrochemical reaction. Fig.9 and 10 shows the oxygen distribution and oxygen reaction rate along the *x* direction, respectively. Compared NPMC CL and Pt CL, it can be seen that although the total reaction surface area of NPMC CL is seven times larger than that of Pt CL, the reaction rate within the NPMC CL is much lower than that in Pt CL, due to much lower $i_0$. As discussed in Section 2, the catalyst loading for the NPMC CL studied in the present study is only 0.283 mg cm$^{-2}$, while in literature it is reported that 1-5 mg cm$^{-2}$ loading can be applied. The higher loading can be achieved by adopting much thicker CLs in practice. Certainly, a much thicker NPMC can further utilize the unconsumed oxygen shown in Fig. 9 and increase the current density. However, a much thicker NPMC will lead to high mass transport resistance for oxygen to arrive at the reactive sites near the CL/PEM interface, especially under high current density [4].

An important implementation for developing PANI-FeCo-C catalysts is acid leaching, which can remove inactive surface species, leading to improved active-site densities and increased micropores (with pore size of few nanometers) [10]. Dominated diffusion mechanism in such micro-pores is Knudsen diffusion, which is much slower than the bulk diffusion in meso or macro-pores (~few tens of nanometers) [9, 22]. In the present study, micropores and mesopores are artificially generated within the solid phase using QSGS to investigate their effects on transport processes and electrochemical reactions. Note that these pores are only generated inside the solid phase, while the shell of the solid phase is not changed, thus guaranteeing the contact between solid phase and ionomer. To focusing on the effects of these pores on mass transport, the distributions of ionomer are not changed, meaning that there is no ionomer within these newly generated pores where thus electrochemical reactions are not allowed. Two structures are generated, namely NPMC A1 and NPMC A2, and their PSD are shown in Fig. 11. For NPMC A1, pores with diameter of 5~10 nm are increased while for NPMC A2, pores with diameter 15-25 nm become more, as can be observed from the PSD. The effective diffusivity for NPMC A1 and NPMC A2 $D_{\text{eff}}/D_b$ are $6.76\times10^{-3}$ and $1.98\times10^{-2}$, respectively. Compared with the original NPMC with $D_{\text{eff}}/D_b = 6.45\times10^{-3}$, micropores in NPMC A1 contributes little to the mass transport; in contrast, NPMC A2 greatly reduces the mass transport resistance and is expected to improve the cell performance. As shown in Fig. 9 and 10, for NPMC A1, the oxygen concentration at the right outlet only increases 21%, and the oxygen reaction rate is slightly increased. For NPMC A2, the oxygen concentration at the outlet increases three times, and the reaction rate enhances remarkably. This confirms that due to less efficient Knudsen diffusion within micropores, mesopores or macropores should be generated for the purpose of reduce the mass transport limitation in NPMC CL, which is in consistent with the experimental results in Ref. [9].

## 6. Conclusion

In this study, nanoscale structures of CL are reconstructed using a reconstructed method called QSGS [41]. The reconstruction processes follows the experimental fabrication process including carbon seed generation, carbon phase growth, Pt deposition and Nafion coverage [23]. Detailed distributions of Pt, carbon, ionomer and viod space are obtained from the reconstruction processes due to the high resolution of the reconstructed structures. Characterization analyses of the reconstructed structures are implemented including pore size distribution, specific area and phase connectivity. The 13 direction averaging method proposed in Ref. [22] is employed to calculate the pore size. All the pore sizes are located in the range of 5-130 nm, indicating the nanoscale characteristics of the CL. The maximum of the PSD is located at around 50 nm, in good agreement with the experimental results [43]. With resolution of 5nm, the specific surface area of the reconstructed CL is 86.2 $m^2$ $g^{-1}$, which is relatively lower than the experimental results in [43], possibly due to failing capture of pores with size smaller than 5nm [43]. A connected phase labeling algorithm is developed to identify the "transport" and "dead" portions of certain phase [44]. For C/Pt phase, void space and ionomer, the volume fraction of the "transport" portion is 99.80%, 99.69% and 99.86%, respectively, in good agreement with the experimental results in [13].

Macroscopic transport properties of the reconstructed CL are numerically predicted using self-developed model based on the LBM. For the effective diffusivity without considering Knudsen diffusion, the values predicted in the present study are much lower than that of Lange et al. [22]. The reason is the the nonuniform ionomer distributions in the reconstructed CL, which greatly increase the tortursity of the path for oxygen transport in the void space, thus obviously reducing the effective diffusivity. The tortuosity predicted is much higher than that commonly used in Bruggeman equation. Based on Eq. (28) and assuming a power relationship between ε and τ ( $\tau = \varepsilon^{\alpha}$ ), the averaged $\alpha$ predicted in the present work is -1.48, while that of Lang et al. [22] is -

0.74 and that predicted very recently by Wu and Jiang is much lower as -2.11 [31], all of which are higher than -0.5 used in Bruggeman equation. In addition, Knudsen diffusion plays a significant role in oxygen diffusion and significantly reduces the effective diffusivity. Detailed structures of the carbon black as well as the distribution characteristics of ionomer greatly affect the effective diffusivity. The effective proton conductivity predicted is close the results in [25], much lower than experimental results [51], indicating other proton transport mechanisms such as proton transport in liquid water [52].

Reactive transport in CL of PEMFCs with NPMC CL and Pt CL are numerically investigated at the nanoscale using the LBM. For the first step study, the porous structures of the NPMC CL are assumed to the same as that of Pt CL, a assumption which is believed to be reasonable based on the SEM images of of the PANI-FeCo-C catalysts studied in the present study [10] and that of Pt CL [23]. The entire surface of the solid phase in PANI-FeCo-C CL is reactive, while in Pt CL only sites with Pt nanoparticles are reactive. The simulation results find out that although the total reaction surface area of NPMC CL is seven times larger than that of Pt CL, the reaction rate within the NPMC CL is much lower than that in Pt CL, due to much lower reaction rate. Much high loading of NPMC can be applied. In addition, micropores and mesopores are generated in the solid phase of the NPMC CL. The numerical simulation results of such modified CL find that micropores contribute little to enhance the mass transport. Mesopores (few tens of nanometers) or macropores are required to increase the mass transport rate.

**Acknowledgement**

The authors acknowledge the support of LANL's LDRD Program, Institutional Computing Program, National Nature Science Foundation of China (No. 51136004), and NNSFC international-joint key project (No. 51320105004). Li Chen appreciates the helpful discussions with Doctor Yutong Mu from Xi'an Jiaotong University, China, Doctor WenZhen Fang from Xi'an Jiaotong University, China, and Prof. Gregory A. Voth from the University of Chicago, USA.

**Reference**

[1] H.A. Gasteiger, J.E. Panels, S.G. Yan, Dependence of PEM fuel cell performance on catalyst loading, Journal of Power Sources, 127(1–2) (2004) 162-171.
[2] H.A. Gasteiger, S.S. Kocha, B. Sompalli, F.T. Wagner, Activity benchmarks and requirements for Pt, Pt-alloy, and non-Pt oxygen reduction catalysts for PEMFCs, Applied Catalysis B: Environmental, 56(1–2) (2005) 9-35.
[3] B.D. James, J.A. Kalinoski, Mass production cost estimation for direct $H_2$ PEM fuel cell system for automotive applications, in: U.S. Department of Energy-Hydrogen Program, 2008 Annual Progress Report, 2008.
[4] F. Jaouen, E. Proietti, M. Lefèvre, R. Chenitz, J.-P. Dodelet, G. Wu, H.T. Chung, C.M. Johnston, P. Zelenay, Recent advances in non-precious metal catalysis for oxygen-reduction reaction in polymer electrolyte fuel cells, Energy Environ. Sci., 4 (2011) 114-130.
[5] D.C. Higgins, Z. Chen, Recent progress in non-precious metal catalysts for PEM fuel cell applications, The Canadian Journal of Chemical Engineering, 91(12) (2013) 1881-1895.
[6] R. Jasinski, A new fuel cell cathode catalyst, Nature, 201(4925) (1964) 1212-1213.
[7] H.A. Gasteiger, N.M. Marković, Just a dream—or future reality?, Science, 324(5923) (2009) 48-49.
[8] M. Lefèvre, E. Proietti, F. Jaouen, J.-P. Dodelet, Iron-based catalysts with improved oxygen reduction activity in polymer electrolyte fuel cells, Science, 324(5923) (2009) 71-74.
[9] E. Proietti, F. Jaouen, M. Lefèvre, N. Larouche, J. Tian, J. Herranz, J.-P. Dodelet, Iron-based cathode catalyst with enhanced power density in polymer electrolyte membrane fuel cells, Nat Commun, 2 (2011) 416.
[10] G. Wu, K.L. More, C.M. Johnston, P. Zelenay, High-performance electrocatalysts for oxygen reduction derived from polyaniline, iron, and cobalt, Science, 332(443-447) (2011).
[11] S. Gupta, D. Tryk, I. Bae, W. Aldred, E. Yeager, Heat-treated polyacrylonitrile-based catalysts for oxygen electroreduction, J Appl Electrochem, 19(1) (1989) 19-27.
[12] S. Litster, G. McLean, PEM fuel cell electrodes, Journal of Power Sources, 130(1–2) (2004) 61-76.
[13] S. Thiele, T. Fürstenhaupt, D. Banham, T. Hutzenlaub, V. Birss, C. Ziegler, R. Zengerle, Multiscale tomography of nanoporous carbon-supported noble metal catalyst layers, Journal of Power Sources, 228(0) (2013) 185-192.
[14] V. Nallathambi, J.-W. Lee, S.P. Kumaraguru, G. Wu, B.N. Popov, Development of high performance carbon composite catalyst for oxygen reduction reaction in PEM Proton Exchange Membrane fuel cells, Journal of Power Sources, 183(1) (2008) 34-42.
[15] F. Jaouen, J.-P. Dodelet, Average turn-over frequency of O2 electro-reduction for Fe/N/C and Co/N/C catalysts in PEFCs, Electrochimica Acta, 52(19) (2007) 5975-5984.
[16] T. Berning, D.M. Lu, N. Djilali, Three-dimensional computational analysis of transport phenomena in a PEM fuel cell, Journal of Power Sources, 106(1–2) (2002) 284-294.
[17] W.Q. Tao, C.H. Min, X.L. Liu, Y.L. He, B.H. Yin, W. Jiang, Parameter sensitivity examination and discussion of PEM fuel cell simulation model validation: Part I. Current status of modeling research and model development, Journal of Power Sources, 160(1) (2006) 359-373.
[18] W. Sun, B.A. Peppley, K. Karan, An improved two-dimensional agglomerate cathode model to study the influence of catalyst layer structural parameters, Electrochimica Acta, 50(16–17) (2005) 3359-3374.
[19] G. Wang, P.P. Mukherjee, C.-Y. Wang, Direct numerical simulation (DNS) modeling of PEFC electrodes: Part II. Random microstructure, Electrochimica Acta, 51(15) (2006) 3151-3160.
[20] G. Wang, P.P. Mukherjee, C.-Y. Wang, Optimization of polymer electrolyte fuel cell cathode catalyst layers via direct numerical simulation modeling, Electrochimica Acta, 52(22) (2007) 6367-6377.
[21] S.H. Kim, H. Pitsch, Reconstruction and effective transport properties of the catalyst layer in PEM fuel cells, Journal of The Electrochemical Society, 156(6) (2009) B673-B681.

[22] K.J. Lange, P.-C. Suia, N. Djilali, Pore scale simulation of transport and electrochemical reactions in reconstructed PEMFC catalyst layers, Journal of the Electrochemical Society, 157(10) (2010) B1434-B1442.
[23] N.A. Siddique, F. Liu, Process based reconstruction and simulation of a three-dimensional fuel cell catalyst layer, Electrochimica Acta, 55(19) (2010) 5357-5366.
[24] R. Barbosa, J. Andaverde, B. Escobar, U. Cano, Stochastic reconstruction and a scaling method to determine effective transport coefficients of a proton exchange membrane fuel cell catalyst layer, Journal of Power Sources, 196(3) (2011) 1248-1257.
[25] K.J. Lange, P.-C. Sui, N. Djilali, Pore scale modeling of a proton exchange membrane fuel cell catalyst layer: Effects of water vapor and temperature, Journal of Power Sources, 196(6) (2011) 3195-3203.
[26] K.J. Lange, P.-C. Sui, N. Djilali, Determination of effective transport properties in a PEMFC catalyst layer using different reconstruction algorithms, Journal of Power Sources, 208(0) (2012) 354-365.
[27] P.P. Mukherjee, C.-Y. Wang, Stochastic microstructure reconstruction and direct numerical simulation of the PEFC catalyst layer, Journal of The Electrochemical Society, 153(5) (2006) A840-A849.
[28] L. Chen, H.-B. Luan, Y.-L. He, W.-Q. Tao, Pore-scale flow and mass transport in gas diffusion layer of proton exchange membrane fuel cell with interdigitated flow fields, International Journal of Thermal Sciences, 51 (2012) 132-144.
[29] J. Park, M. Matsubara, X. Li, Application of lattice Boltzmann method to a micro-scale flow simulation in the porous electrode of a PEM fuel cell, Journal of Power Sources, 173(1) (2007) 404-414.
[30] P.P. Mukherjee, C.Y. Wang, Q. Kang, Mesoscopic modeling of two-phase behavior and flooding phenomena in polymer electrolyte fuel cells, Electrochimica Acta, 54(27) (2009) 6861-6875.
[31] W. Wu, F. Jiang, Microstructure reconstruction and characterization of PEMFC electrodes, online, International Journal of Hydrogen Energy.
[32] J. Zhang, W. Yang, L. Xu, Y. Wang, Simulation of the catalyst layer in PEMFC based on a novel two-phase lattice model, Electrochimica Acta, 56(20) (2011) 6912-6918.
[33] S.Y. Chen, G.D. Doolen, Lattice Boltzmann methode for fluid flows, Annual Review of Fluid Mechanics, 30 (1998) 329-364.
[34] C.K. Aidun, J.R. Clausen, Lattice-Boltzmann method for complex flows, Annual Review of Fluid Mechanics, 42(439-472) (2010).
[35] L. Chen, Q. Kang, Y.-L. He, W.-Q. Tao, Mesoscopic study of the effects of gel concentration and materials on the formation of precipitation patterns, Langmuir, 28(32) (2012) 11745–11754.
[36] L. Chen, Q. Kang, B.A. Robinson, Y.-L. He, W.-Q. Tao, Pore-scale modeling of multiphase reactive transport with phase transitions and dissolution-precipitation processes in closed systems, Physical Review E, 87(4) (2013) 043306.
[37] Q. Kang, D. Zhang, S. Chen, Simulation of dissolution and precipitation in porous media, Journal of Geophyiscal Research, 108 (2003) B10, 2505.
[38] Q. Kang, D. Zhang, S. Chen, X. He, Lattice Boltzmann simulation of chemical dissolution in porous media, Physical Review E, 65(3) (2002) 036318.
[39] L. Chen, Q. Kang, Y. Mu, Y.-L. He, W.-Q. Tao, A critical review of the pseudopotential multiphase lattice Boltzmann model: Methods and applications, International Journal of Heat and Mass Transfer, 76(0) (2014) 210-236.
[40] S. Thiele, R. Zengerle, C. Ziegler, Nano-morphology of a polymer electrolyte fuel cell catalyst layer—imaging, reconstruction and analysis, Nano Res., 4(9) (2011) 849-860.
[41] M. Wang, J. Wang, N. Pan, S. Chen, Mesoscopic predictions of the effective thermal conductivity for microscale random porous media, Phys. Rev. E, 75(036702) (2007).
[42] C.V. Rao, B. Viswanathan, Monodispersed platinum nanoparticle supported carbon electrodes for hydrogen oxidation and oxygen reduction in proton exchange membrane fuel cells, The Journal of Physical Chemistry C, 114(18) (2010) 8661-8667.

[43] T. Soboleva, X. Zhao, K. Malek, Z. Xie, T. Navessin, S. Holdcroft, On the micro-, meso-, and macroporous structures of polymer electrolyte membrane fuel cell catalyst layers, ACS Applied Materials & Interfaces, 2(2) (2010) 375-384.
[44] C. Pan, E. Dalla, D. Franzosi, C.T. Miller, Pore-scale simulation of entrapped non-aqueous phase liquid dissolution, Advances in Water Resources, 30(3) (2007) 623-640.
[45] E.L. Cussler, Diffusion: mass transfer in fluid systems, third edition, Cambridge University, New York, 1997.
[46] R.E. Cunningham, R.J.J. Williams, Diffusion in gases and porous media, Plenum, New York, 1980.
[47] W.G. Pollard, R.D. Present, On gaseous self-diffusion in straight cylindrical pores, Phys. Rev. , 73 (1948 ) 762-774.
[48] D.M. Bernardi, M.W. Verbrugge, A mathematical model of the solid-polymer-electrolyte fuel cell, J. Electrochem. Soc., 139(9) (1992) 2477–2491.
[49] L. Chen, Y.-L. Feng, C.-X. Song, L. Chen, Y.-L. He, W.-Q. Tao, Multi-scale modeling of proton exchange membrane fuel cell by coupling finite volume method and lattice Boltzmann method, International Journal of Heat and Mass Transfer, 63(0) (2013) 268-283.
[50] Z. Yu, R.N. Carter, Measurement of effective oxygen diffusivity in electrodes for proton exchange membrane fuel cells, Journal of Power Sources, 195(4) (2010) 1079-1084.
[51] Y. Liu, M.W. Murphy, D.R. Baker, W. Gu, C. Ji, J. Jorne, H.A. Gasteiger, Proton conduction and oxygen reduction kinetics in PEM fuel cell cathodes: effects of ionomer-to-carbon ratio and relative humidity, Journal of The Electrochemical Society, 156(8) (2009) B970-B980.
[52] P. Choi, N.H. Jalani, R. Datta, Thermodynamics and proton transport in nafion: II. proton diffusion mechanisms and conductivity, Journal of The Electrochemical Society, 152(3) (2005) E123-E130.

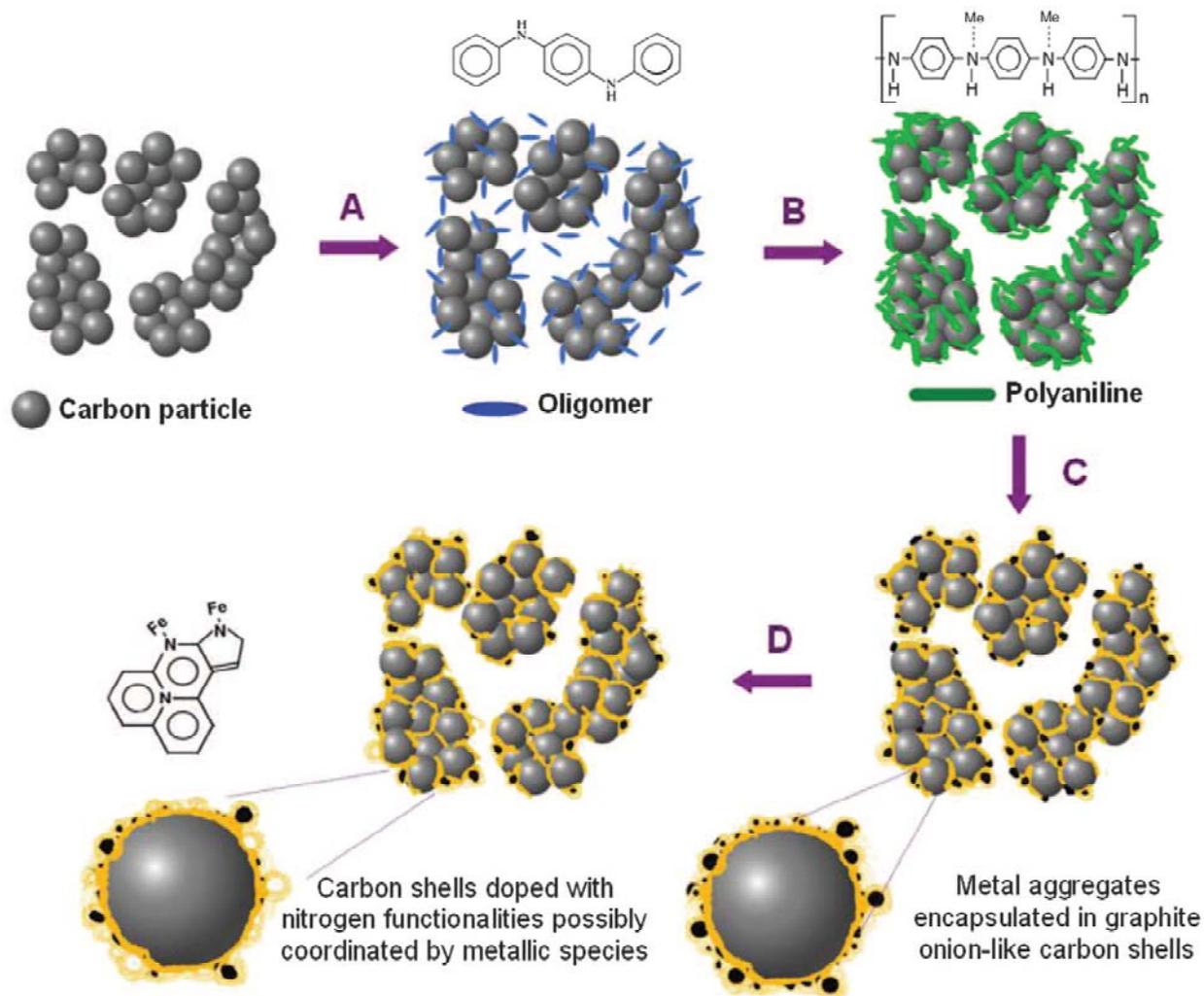

Fig.1 Schematic diagram of the synthesis of PANI-M-C catalysts. (a) Mixing of high–surface area carbon with aniline oligomers and transition-metal precursor (M: Fe and/or Co). (b) Oxidative polymerization of aniline by addition of APS. (c) First heat treatment in N2 atmosphere. (d) Acid leaching. The second heat treatment after acid leach is not shown. (Reproduced from Fig. 1 in Ref. [10])

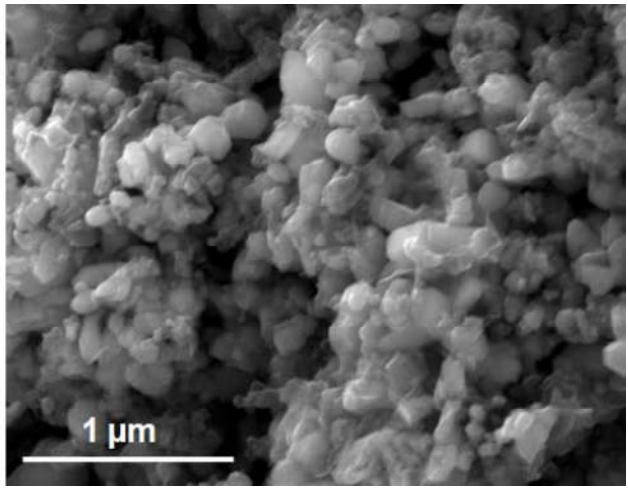

Fig.2 SEM of the PANI-FeCo-C with heat treatment of 900°C (Reproduced from Fig. S6 in Ref. [10])

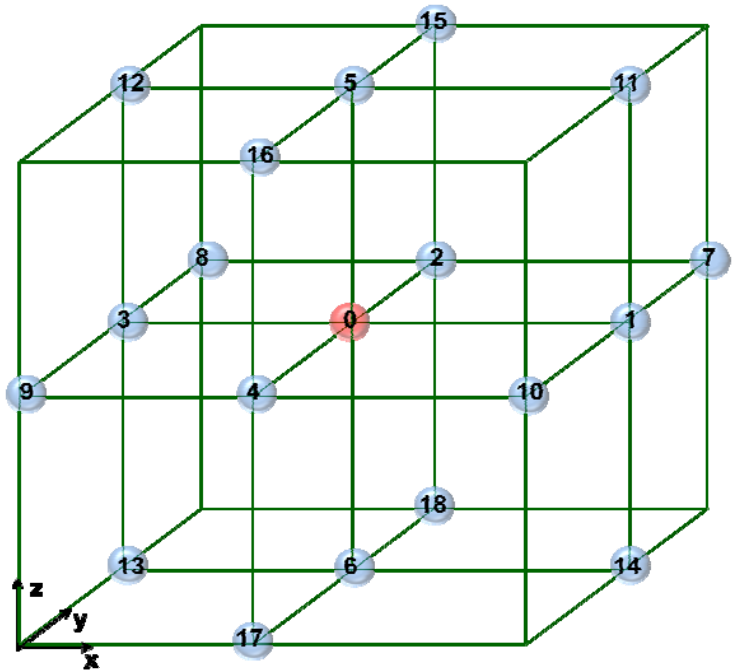

Fig. 3 6 directions for growth in the QSGS or the D3Q19 lattice model in LBM

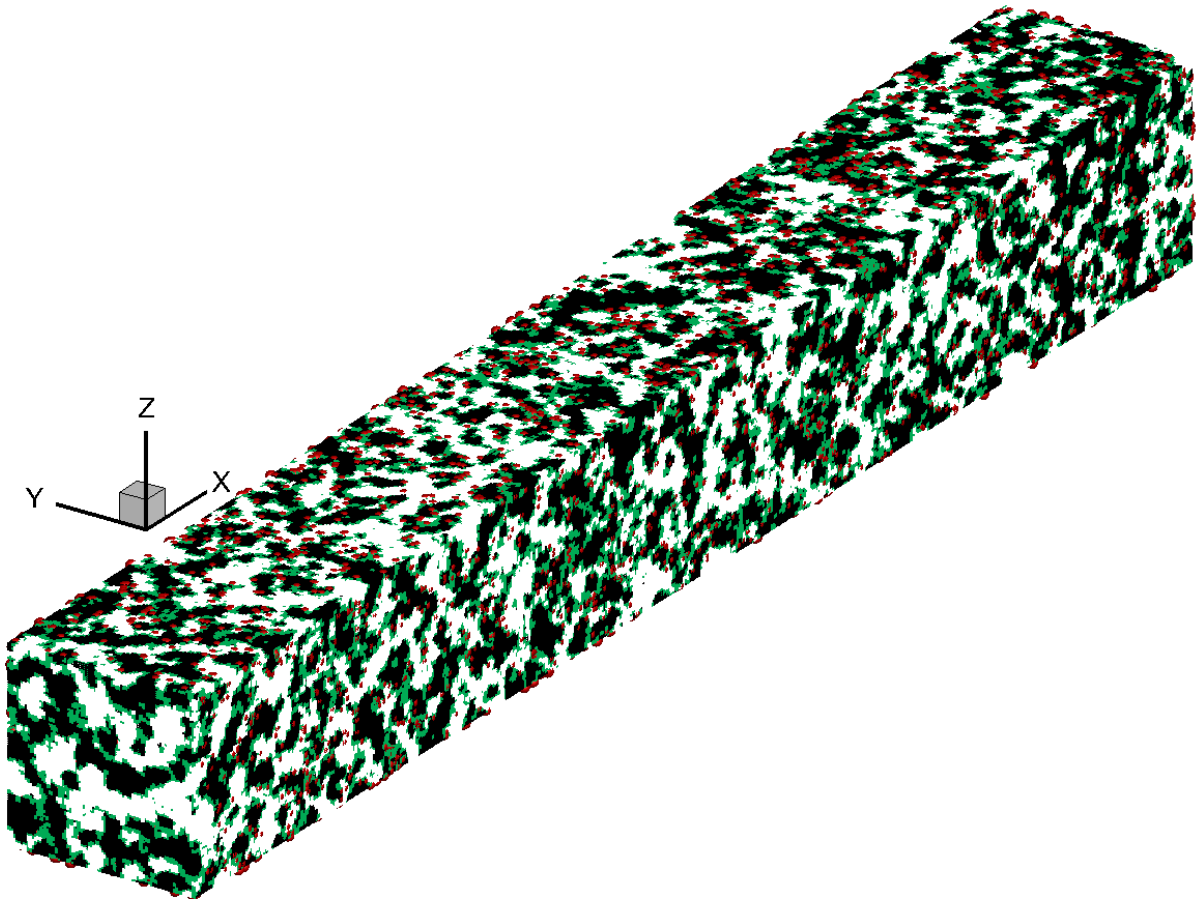

(a) 3D structures

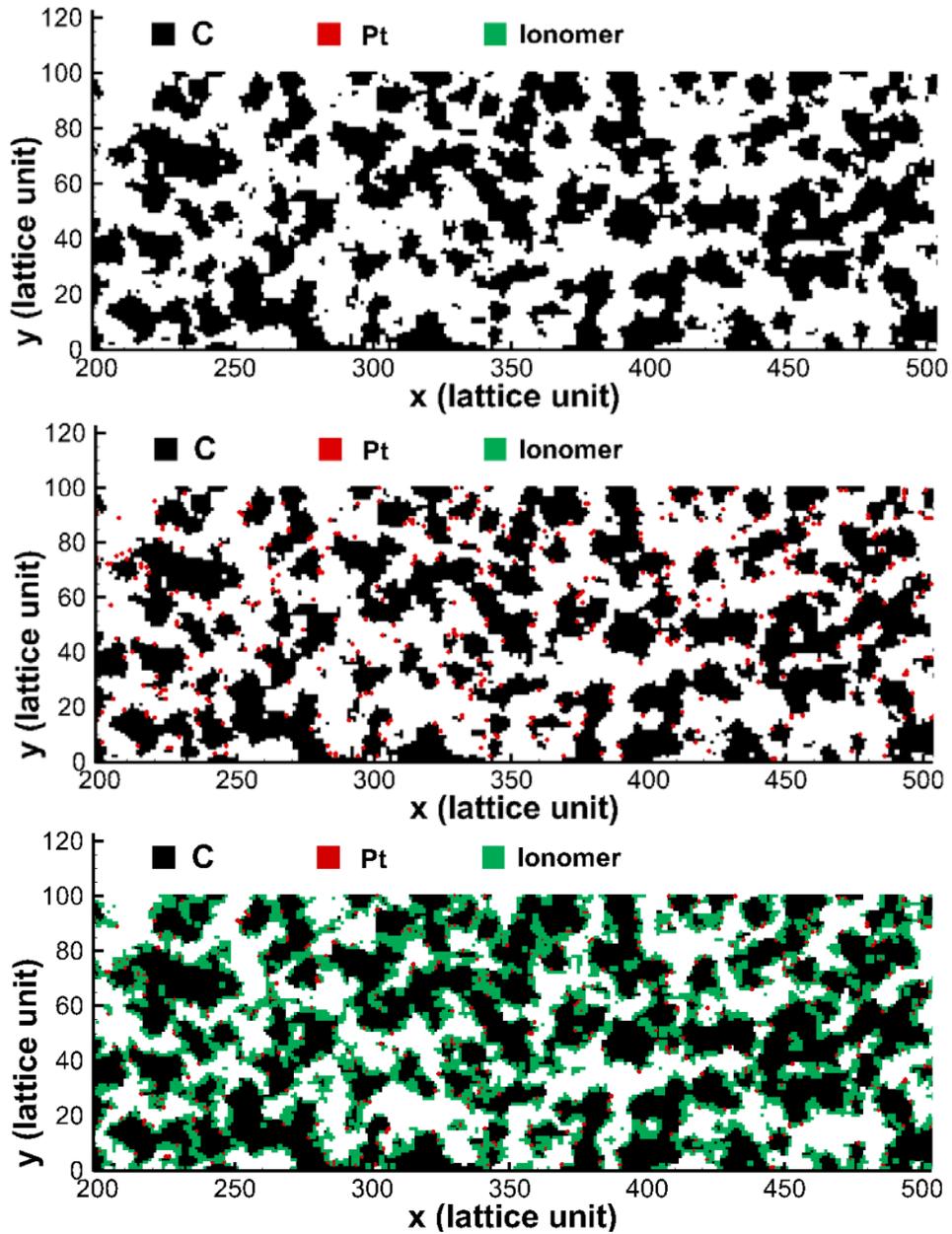

(b) Detailed distributions of carbon black (top), C/Pt (middle) and Nafion covered on C/Pt (bottom)

Fig.4 Reconstructed nanoscale structures of CCL

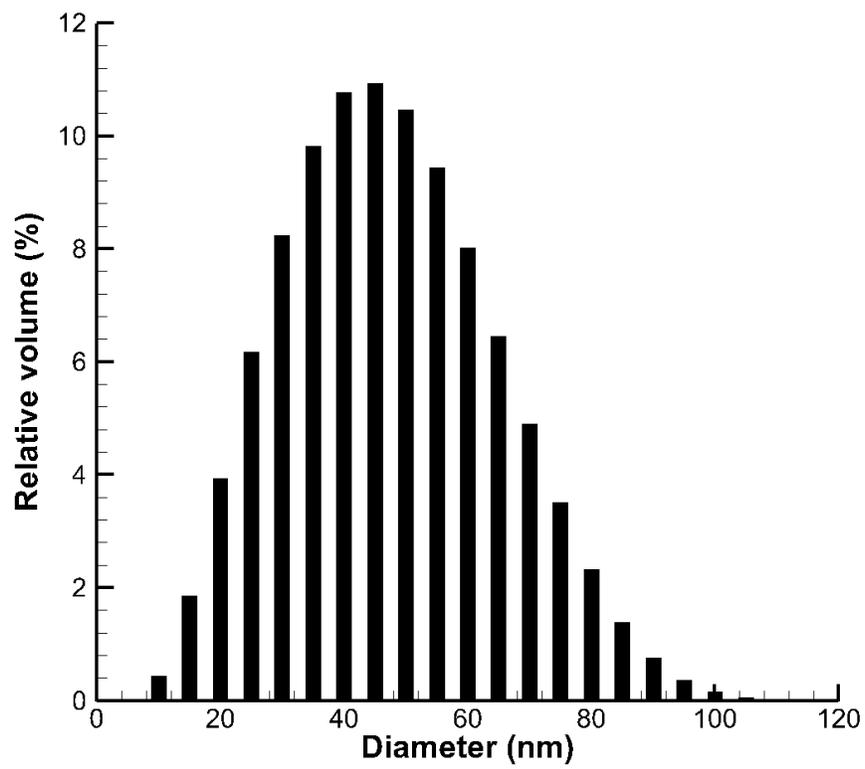

Fig.5 Pore size distribution of the reconstructed CL structure

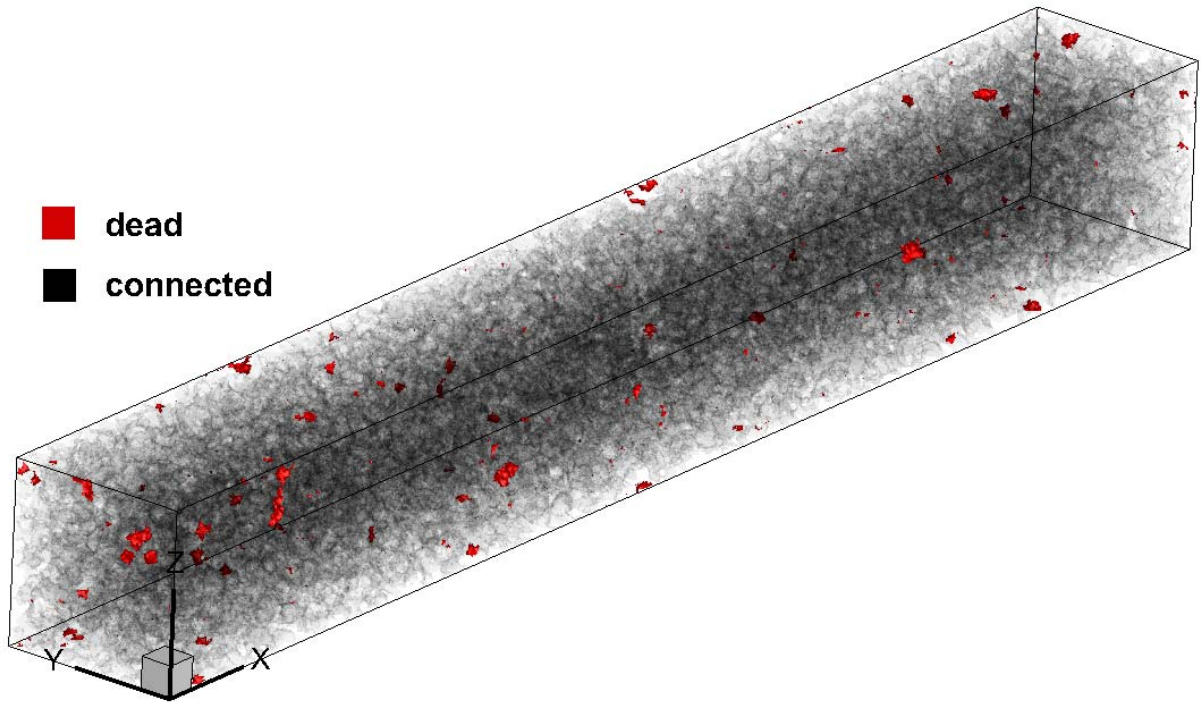

Fig. 6 The "transport" and "dead" portions of solid phase in the reconstructed CL

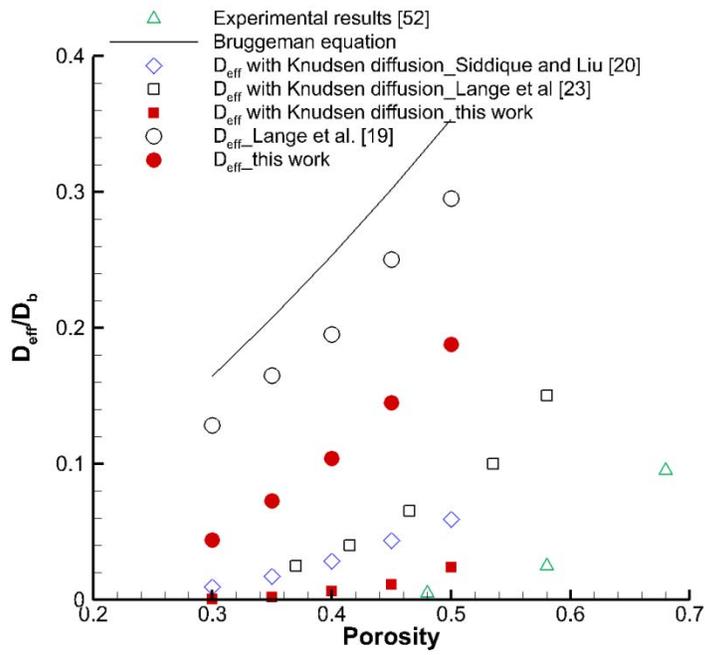

(a) Effective diffusivity under different porosity

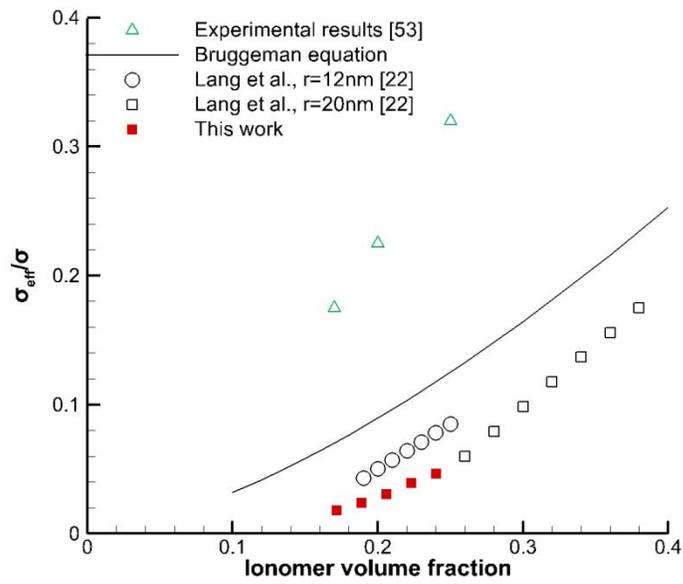

(b) Effective proton conductivity under different ionomer volume fraction

Fig.7 Macroscopic transport properties obtained from LBM simulations

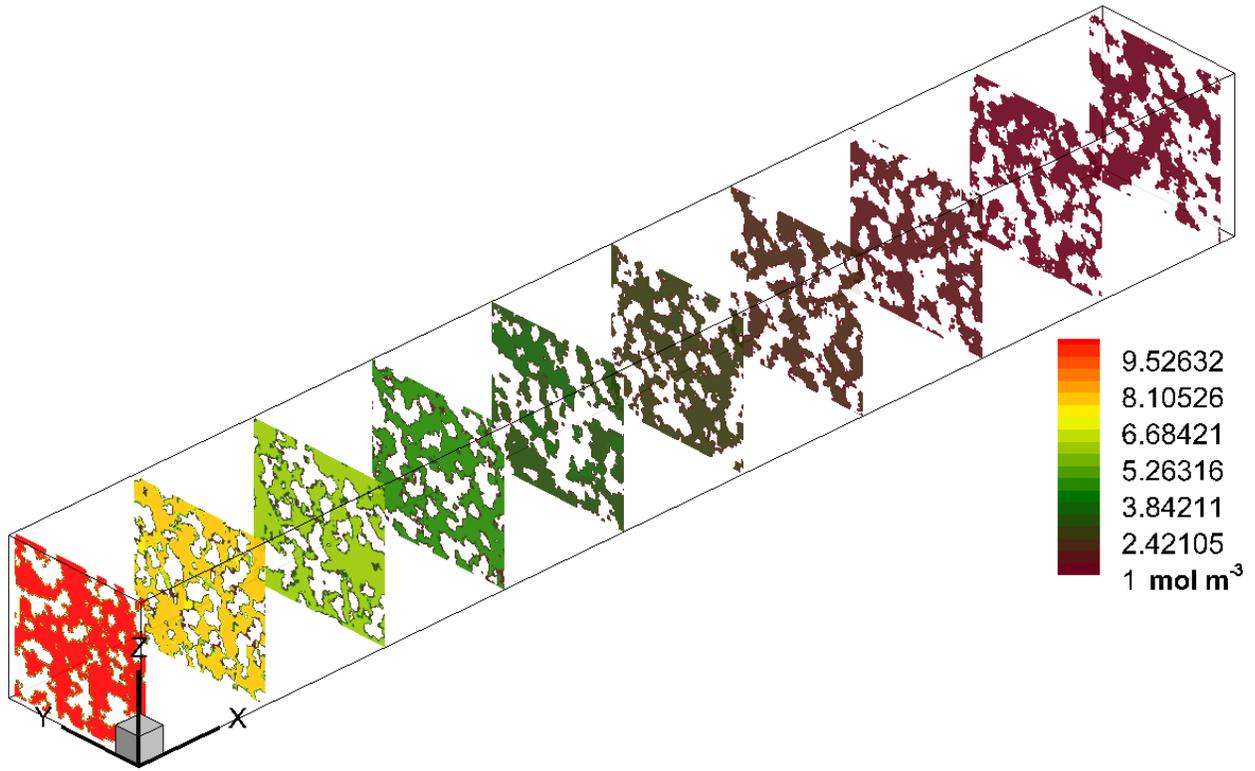

Fig. 8 Oxygen concentration in the NPMC CL

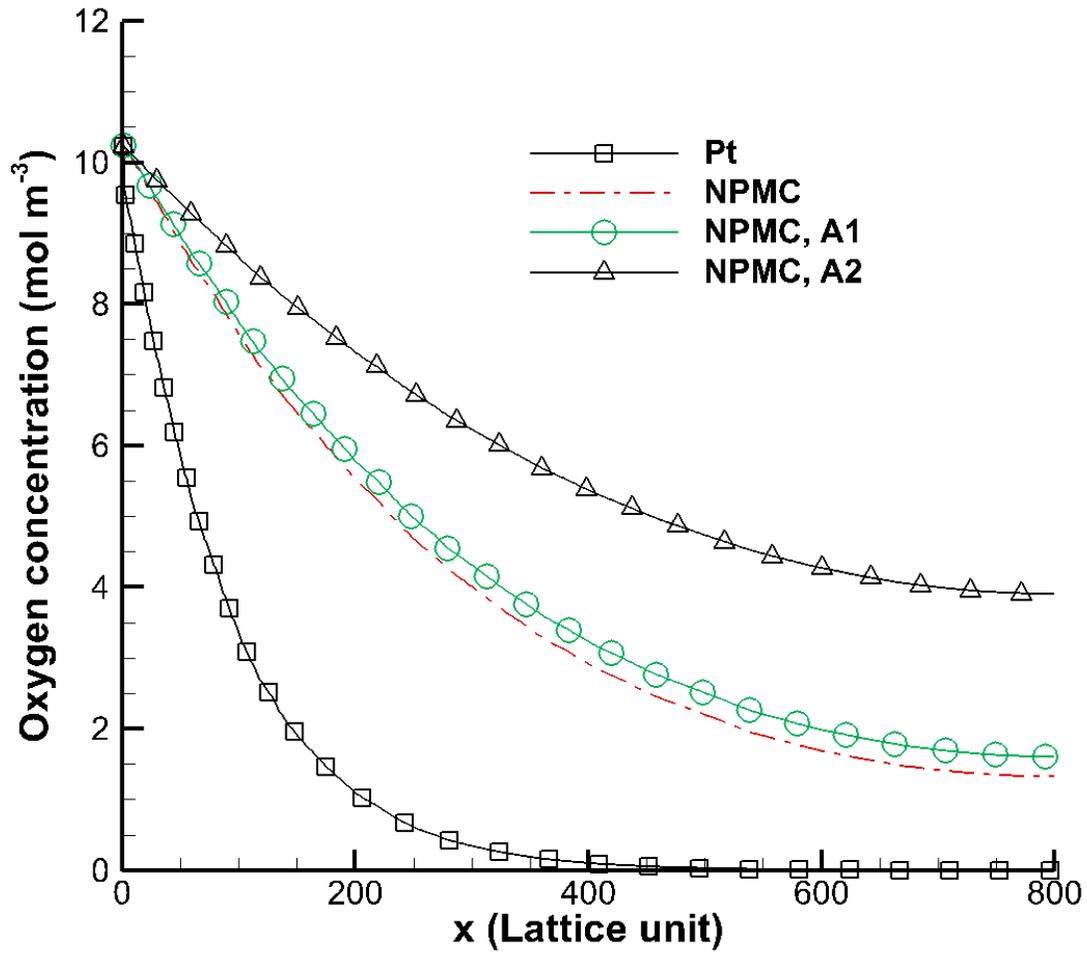

Fig. 9 Distributions of oxygen distributions along *x* direction

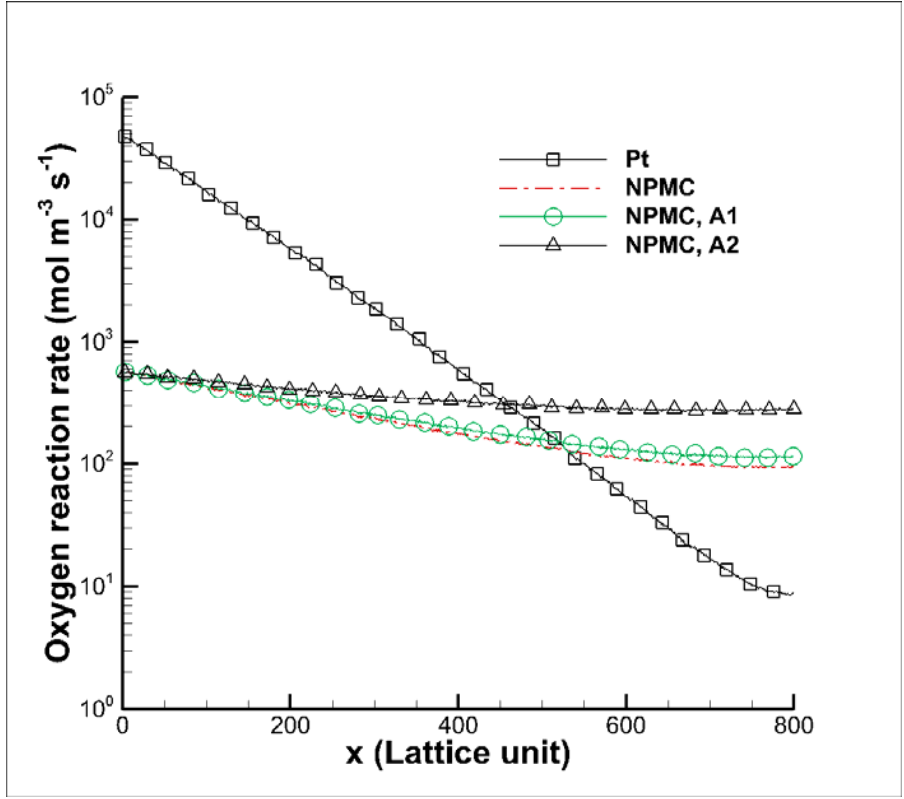

Fig. 10 Distributions of reaction rate along *x* direction

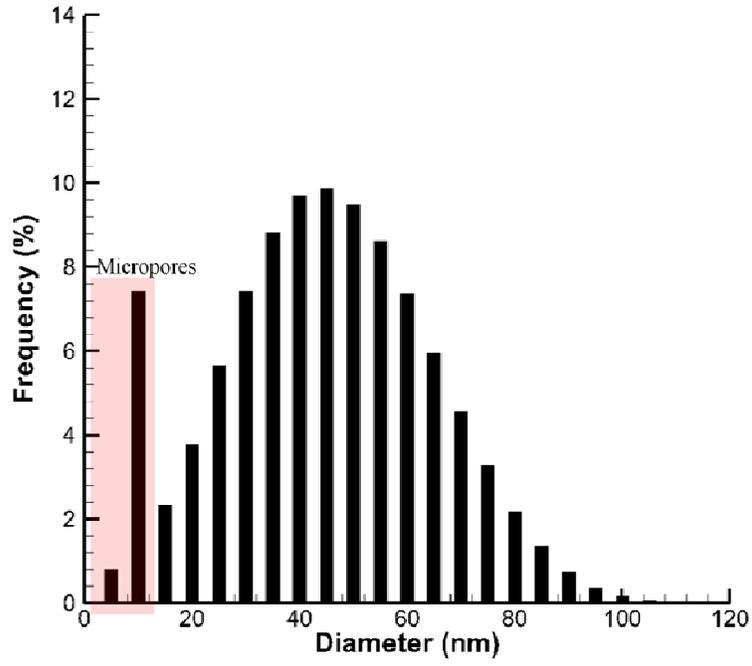

(a) NPMC A1, micropores generated

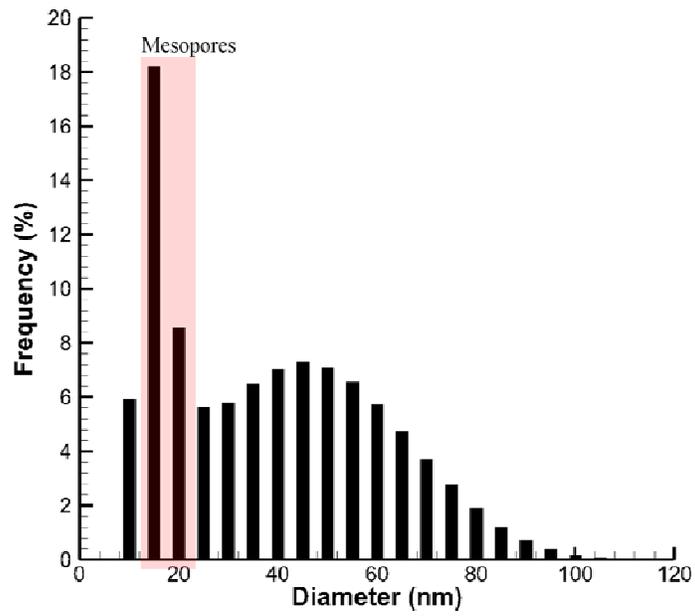

(b) NPMC A1, mesopores generated

Fig. 11 Pore size distribution for NPMC with micropores and mesopores generated

Table 1 Parameter values used in the simulation

| Parameters | | Value |
|---|---|---|
| Temperature | $T$ | 333K |
| Pressure | $P$ | 1.5 bar |
| Reference oxygen concentration | $C_{O2,ref}$ | 40.96 m³ mol⁻¹ |
| cathode transfer coefficient | $\alpha$ | 0.61 |
| modified exchange current density | $i_{0,m}$ | 0.01 A m⁻² |
| Ionomer potential | $\varphi_e$ | 0.6V |
| Oxygen concentration at GDL/CL interface | $C_{O2,ref}$ | 10.28 m³ mol⁻¹ |